\documentclass[a4paper]{amsart}
\usepackage{latexsym}
\usepackage{amsmath}
\usepackage{amssymb}
\usepackage{amsfonts}
\usepackage{esint}
\usepackage{mathrsfs}
\usepackage{graphicx}
\usepackage[all,cmtip]{xy} 
\usepackage{color}
\definecolor{Blue}{rgb}{0.,0.,1.}
\definecolor{Red}{rgb}{0.,0.,0.}
\newcounter{smallarabics}
\newenvironment{arabicenumerate}
{\begin{list}{{\normalfont\textrm{(\arabic{smallarabics})}}}
  {\usecounter{smallarabics}\setlength{\itemindent}{0cm}
   \setlength{\leftmargin}{5ex}\setlength{\labelwidth}{4ex}
   \setlength{\topsep}{0.75\parsep}\setlength{\partopsep}{0ex}
   \setlength{\itemsep}{0ex}}}
{\end{list}}

\newcounter{smallroman}

\newcommand{\ben}{\begin{arabicenumerate}}  
\newcommand{\een}{\end{arabicenumerate}}

\def\init{\setcounter{equation}{0}}

\newtheorem{theoreme}{Theorem}[section]
\newtheorem{hypothesis}{Hypothesis}[section]
\newtheorem{proposition}[theoreme]{Proposition}
\newtheorem{lemma}[theoreme]{Lemma}
\newtheorem{definition}[theoreme]{Definition}

\newtheorem{remark}[theoreme]{Remark}
\newtheorem{example}[theoreme]{Example}
\newcommand{\beq}{\begin{equation}}
\newcommand{\eeq}{\end{equation}}

\newcommand{\bex}{\begin{example}}
\newcommand{\eex}{\end{example}}
\def\bel{\begin{lemma}}
\def\eel{\end{lemma}}
\def\bet{\begin{theoreme}}
\def\eet{\end{theoreme}}
\def\bed{\begin{definition}}
\def\eed{\end{definition}}
\def\ber{\begin{remark}}
\def\eer{\end{remark}}

\def\rr{{\mathbb R}}
\def\zz{{\mathbb Z}}
\def\cc{{\mathbb C}}
\def\nn{{\mathbb N}}

\def\part{{\rm par}}

\def\cpl{{\rm cpl}}

\def\bar{\overline}

\def\cinf{C^\infty}
\def\c0inf{C_0^\infty}

\def\proof{
\noindent{\bf Proof.}\ \ }

\def\h{{\rm h}}

\def\tg{{\rm tg}}

\def\esssupp{{\rm ess\,supp}}

\def\wed{{\odot}}

\def\cY{{\mathcal Y}}

\def\cD{{\mathcal D}}

\def\CCR{{\rm CCR}}

\let\isanspoint\i
\let\i\relax
\def\i{{\rm i}}

\def\ad{{\rm ad}}

\def\Dom{{\rm Dom}}

\def\sgn{{\rm sgn}}

\def\Ker{{\rm Ker}}

\def\qed{$\Box$\medskip}

\def \p{ \partial}

\def\12{\frac{1}{2}}

\def\ad{{\rm ad}}
\def\e{{\rm e}}

\def\Ran{{\rm Ran}}
\def\tS{\tilde{S}}

\def\bbbone{{\mathchoice {\rm 1\mskip-4mu l} {\rm 1\mskip-4mu l}
{\rm 1\mskip-4.5mu l} {\rm 1\mskip-5mu l}}}
\def\one{\bbbone}
\def\cH{{\mathcal H}}

\def\sgn{{\rm sgn}}

\def\Ker{{\rm Ker}}

\def\coinf{C_0^\infty}

\def\C{{\mathcal C}}

\def \p{ \partial}

\def\12{\frac{1}{2}}

\DeclareMathOperator\supp{supp}

\def\ad{{\rm ad}}
\def\e{{\rm e}}

\def\Ran{{\rm Ran}}
\def\tS{\tilde{S}}

\def\cH{{\mathcal H}}

\def\bep{\begin{proposition}}
\def\eep{\end{proposition}}

\def\Op{{\rm Op}}

\newcommand{\mat}[4]{\left(\begin{array}{cc}#1 &#2  \\ #3 &#4 \end{array}\right)}

\makeatletter
\newcommand*{\defeq}{\mathrel{\rlap{%
                     \raisebox{0.3ex}{$\m@th\cdot$}}%
                     \raisebox{-0.3ex}{$\m@th\cdot$}}%
                     =}
\makeatother
\makeatletter
\newcommand*{\eqdef}{=\mathrel{\rlap{%
                     \raisebox{0.3ex}{$\m@th\cdot$}}%
                     \raisebox{-0.3ex}{$\m@th\cdot$}}%
                     }
\makeatother

\def\CARal{{\rm C\hskip 0.25 em \hbox{\raise 1.72 ex 
\hbox{$\scriptscriptstyle\rm al$}\kern -0.57 em A}R}}

\def\otimesal{\mathop{\hbox{\raise 1.5 ex
  \hbox{$\scriptscriptstyle\rm al$}
\kern -0.92 em \hbox{$\otimes$}}}}
\def\oplusal{\mathop{\hbox{\raise 1.5 ex
  \hbox{$\scriptscriptstyle\rm al$}
\kern -0.92 em \hbox{$\oplus$}}}}
\def\Gammal{\hbox{\raise 1.68 ex 
\hbox{$\scriptscriptstyle\rm al$}\kern -0.50 em $\Gamma$}}
\def\Bal{\hbox{\raise 1.68 ex 
\hbox{$\scriptscriptstyle\rm  al$}\kern -0.50 em $B$}}
\def\CARal{{\rm C\hskip 0.25 em \hbox{\raise 1.72 ex 
\hbox{$\scriptscriptstyle\rm al$}\kern -0.57 em A}R}}

\def\cE{{\mathcal E}}
\begin{document}
\def\forms{\raisebox{0.3ex}{$\scriptstyle\bigwedge$}}
\def\wed{\wedge}
\def\rx{{\rm x}}
\def\WF{{\rm WF}}
\newcommand{\co}[1]{T^{*}\! #1}
\newcommand{\coo}[1]{T^{*}\! #1\backslash Z}
\def\cN{\mathcal{N}}
\def\Sol{{\rm Sol}_{\rm sc}}
\def\bS{\mathbb{S}^{d-1}}
\def\musc{(\mu\rm{sc})}
\def\ts{\tilde{s}}
\def\ttet{\tilde{\theta}}
\newcommand{\tra}[1]{\mskip-6mu\upharpoonright_{#1}\mskip+4mu}
\newcommand{\traa}[1]{\mskip-6mu\upharpoonright_{#1}}
\newcommand{\traaa}[1]{\mskip-6mu\upharpoonright_{#1}\mskip-4mu}

\newcommand{\Red}{\color{black}}

\title[Construction of Hadamard states by  characteristic Cauchy problem]{ Construction of Hadamard states by  characteristic Cauchy problem}

\author{  C. G\'erard }
\address{D\'epartement de Math\'ematiques, Universit\'e Paris-Sud XI, 91405 Orsay Cedex, France}
\email{christian.gerard@math.u-psud.fr}
\author{ M. Wrochna}
\address{Universit\'e Joseph Fourier (Grenoble 1), Institut Fourier, UMR 5582 CNRS, BP 74 38402 Saint-Martin
d'H\`eres Cedex, France}
\email{michal.wrochna@ujf-grenoble.fr}
\keywords{Hadamard states, microlocal spectrum condition, pseudo-differential calculus, characteristic Cauchy problem, curved spacetimes}
\subjclass[2010]{81T13, 81T20, 35S05, 35S35}
\begin{abstract}
We construct Hadamard states for Klein-Gordon fields in a spacetime $M_{0}$ equal to the interior of the future lightcone $C$ from a base point $p$ in a globally hyperbolic spacetime $(M, g)$. Under some regularity conditions at future infinity of $C$, we identify a boundary symplectic space of functions on $C$, which allows to construct states for Klein-Gordon quantum fields in $M_{0}$ from states on the $\CCR$ algebra associated to the boundary symplectic space. We  formulate the natural microlocal condition on the boundary  state  on $C$ ensuring that the   bulk state it induces in $M_{0}$ satisfies the Hadamard condition. Using  pseudodifferential calculus  on the cone $C$ we construct a large class of Hadamard  boundary states on the boundary with pseudodifferential covariances, and characterize the pure states among them.  We  then show that  these pure boundary  states  induce pure Hadamard states in $M_{0}$.
\end{abstract}

\maketitle

\section{Introduction}\label{secint}

Hadamard states are  widely accepted  as physically admissible  states for non-interacting quantum fields on a curved spacetime, one of the main reasons being their link with the renormalization of the stress-energy tensor, a basic step in the formulation of semi-classical Einstein equations. Furthermore, they are nowadays considered a necessary ingredient in the perturbative formulation of interacting (non-linear) theories {\Red (cf. recent review articles \cite{KM,HW})}.

For Klein-Gordon fields, the  construction of Hadamard states amounts to finding bi-solutions of the Klein-Gordon equation {\Red (called in this context \emph{two-point functions} and denoted here $\lambda^\pm$)} with a specified wave front set (that is, verifying the \emph{microlocal spectrum condition}) and satisfying additionally a positivity property {\Red \cite{radzikowski}}. 

There exist several methods to construct Hadamard states for Klein-Gordon fields: the first  method relies on the Fulling-Narcowich-Wald deformation argument \cite{FNW}, which reduces the construction of Hadamard states on an arbitrary spacetime to the case of ultrastatic spacetimes, where vacuum or thermal states are  easily shown to be Hadamard states.

The second approach, {worked out in \Red\cite{junker,JS,GW}}, uses pseudodifferential calculus on a fixed Cauchy surface $\Sigma$ in $(M, g)$ and relies on the construction of a parametrix for the Cauchy problem on $\Sigma$. To use pseudodifferential calculus, some restrictions on $\Sigma$ and on the behavior of the metric $g$ at spatial infinity are necessary. On the other hand, the method produces a large classes of rather explicit Hadamard states, whose covariances, expressed in terms of Cauchy data are pseudodifferential operators.

Another method, initiated by Moretti \cite{Mo1,Mo2}  applies to {\em conformal field equations}, like the conformal wave equation, on  an {\em asymptotically flat vacuum spacetime} $(M_{0}, g_{0})$.   By asymptotic flatness, there exists a metric $\tilde{g}_{0}$, conformal to $g_{0}$, and a spacetime $(M, \tilde{g})$ such that  $(M_0, \tilde{g}_{0})$ can be causally embedded as an  open set  in  $(M, \tilde{g})$, with the boundary $C= \p M_{0}$ of $M_{0}$ being {\em null} in $(M, \tilde{g})$. States on the {\em boundary symplectic space}, containing  the traces on $C$ of solutions of the wave equation in $M_{0}$, naturally induce states inside $M_{0}$.

This method has been successfully applied   in \cite{Mo1,Mo2} to construct a distinguished Hadamard state for asymptotically flat vacuum spacetimes with past time infinity and then extended to several  other geometrical situations in \cite{DMP1,DMP2,BJ}. Further results also include generalization to Maxwell fields \cite{DS} and linearized gravity \cite{BDM}.

In the present paper we rework systematically the above strategy {\Red in terms of the associated \emph{characteristic Cauchy problem}} in order to construct a large class of Hadamard states (instead of a preferred single one) and to characterize the pure ones. For the sake of clarity, we do not impose geometrical assumptions on $M_0$ that allow to correctly embed it in a larger spacetime $M$. 

Instead we go the other way around and work in an a priori arbitrary globally hyperbolic spacetime {\Red$(M,g)$}, fix a base point $p$ and consider the interior of the future lightcone 
\[ 
C\defeq \p J^{+}(p)\backslash\{p\}
\] 
as the spacetime $M_0$ of main interest, i.e. $M_0\defeq I^{+}(p)$, {\Red where $I^{+}(p)$ (resp. $J^+(p)$) is the time-like (resp. causal) shadow  of $p$, cf. \cite[Sec. 8.1]{W}.} 

We make the following assumption on the geometry of $C$.

\begin{hypothesis}\label{hyp:main} We assume that there exists $f\in \cinf(M)$ such that:
\[
\begin{array}{rl}
(1)&  C\subset f^{-1}(\{0\}), \ \nabla_{a}f\neq 0\hbox{ on }C,\ \nabla_{a}f(p)=0, \ \nabla_{a}\nabla_{b}f(p)= - 2 g_{ab}(p),\\[2mm]
(2)&\hbox{the vector field }\nabla^{a}f \hbox{ is complete on }C.
\end{array}
\]
\end{hypothesis}

Using Hypothesis \ref{hyp:main} one can  construct coordinates $(f,s,\theta)$ near $C$, such that $C\subset\{f=0\}$ and 
\[
g\tra{C}= -2df ds+ h(s, \theta)d\theta^2,
\]
where $h(s, \theta)d\theta^2$ is a Riemannian metric on $\bS$.

Such choice of coordinates allows one to identify $C$ with $\tilde C\defeq \rr \times \bS$. A natural space of smooth functions on $\tilde C$ is then provided by $\cH(\tilde C)$ --- the intersection of Sobolev spaces of all orders, defined using the standard metric $m(\theta)d\theta^2$ on $\bS$.

{\Red We consider the Klein-Gordon operator $P=-\Box_g+ r(x)$ (with $r(x)\in C^{\infty}(M)$ real-valued) and its restriction on $M_0$, denoted $P_0\defeq P\traa{M_0}$.} The \emph{bulk-to-boundary correspondence} can be expressed in this setup as follows. For an appropriate choice of $\beta(s,\theta)\in C^\infty(M_0)$, the restriction map
\[
\rho \phi\defeq (\beta^{-1}\phi)\traa{C}, \quad \phi\in C^{\infty}_{\rm sc}(M_0)
\]
is a monomorphism\footnote{By monomorphism of symplectic spaces we mean an injective linear map that intertwines the symplectic forms.} between the symplectic space of smooth,  space-compact solutions of $P_0$ (endowed with the usual symplectic form induced by the causal propagator) and $\cH(\tilde{C})$, equipped with the symplectic form
\beq\label{eq:sfc}
 \overline{g}_{1}\sigma_{C}  g_{2}\defeq  \int_{\rr\times\bS} (\p_{s}\overline{g}_{1}g_{2}-\overline{g}_{1} \p_{s}g_{2})|m|^{\12}(\theta) ds d\theta, \ g_{1}, g_{2}\in \cH(\tilde{C}).
\eeq
Thus, a quasi-free state on $(\cH(\tilde{C}),\sigma_C)$ with two-point functions $\lambda^\pm$ induces a unique quasi-free state on the usual symplectic space associated to $P_0$.

\subsection*{Product-type pseudodifferential operators} In \cite{GW} we have constructed Hadamard states whose two-point functions on a Cauchy surface $\Sigma$ are pseudodifferential operators. In the present case, the obvious difference is that on the cone $C$, the coordinate $s$ is distinguished both from the point of view of the microlocal spectrum condition (from now on abbreviated $\musc$) and in the expression (\ref{eq:sfc}) for the symplectic form. This suggests that one should rather consider {\em product-type} pseudodifferential operators $\Psi^{p_1,p_2}(\tilde C)$ with symbols satisfying estimates:
\[
|\p_{s}^{\alpha_{1}}\p_{\sigma}^{\beta_{1}}\p_{\theta}^{\alpha_{2}}\p_{\eta}^{\beta_{2}}a(s,\theta,\sigma,\eta)|\in O(\langle\sigma\rangle^{p_1-|\beta_1|}\langle\eta\rangle^{p_2-|\beta_2|})
\]
in the covariables $\xi=(\sigma,\eta)$ relative to the decomposition $\tilde C = \rr \times\bS$. Actually, to cope with the issue that $\sigma_C$ is defined using an operator $D_s\defeq\i^{-1}\partial_s$ whose spectrum is not separated from $\{0\}$ (analogously to the infrared problem in massless theories), we need to introduce a larger class $\tilde\Psi^{p_1,p_2}(\tilde C)$ that includes some operators whose symbol is discontinuous at $\eta=0$. Namely, we set 
\[
\tilde\Psi^{p_1,p_2}(\tilde C) \defeq \Psi^{p_1,p_2}(\tilde C) + B^{-\infty}\Psi^{p_2}(\tilde C),
\]
where $B^{-\infty}\Psi^{p_2}(\tilde C)$ is the class of pseudodifferential operators of order $p_2$ (in the $\theta$ variables) with values in operators on $\rr$ that infinitely increase Sobolev regularity. Then for instance $|D_s|\otimes \one_\theta\in\tilde\Psi^{1,0}(\tilde C)$ although it is not in the pseudodifferential class $\Psi^{1,0}(\tilde C)$.

\subsection*{Summary of results} Our main results can be summarized as follows. We always assume Hypothesis \ref{hyp:main}. If $E, F$ are topological vector spaces, we write $ T: E\to F$ to mean $T: E\to F$ is linear and continuous.

\begin{enumerate}\itemsep1.5mm 
\item[\textbf{1)}]\label{res:had} For pairs\footnote{We work with charged fields, in which case it is natural to associate a \emph{pair} of two-point functions to a quasi-free state, cf. \ref{ss:qfree}. The charged and neutral approaches are equivalent.} of two-point functions $\lambda^\pm$ on $C$ satisfying $\lambda^\pm : \cH(C)\to\cH(C)$, we give in Thm. \ref{th8.1} conditions on $\WF(\lambda^\pm)$ that guarantee that the corresponding two-point functions on $M_0$ satisfy $\musc$. This is essentially an adaptation of the results of \cite{Mo2} to our framework.
\item[\textbf{2)}]\label{res:constr} In Thm. \ref{thm:constr} we construct a large class of Hadamard states by specifying their two-point functions $\lambda^\pm\in\tilde\Psi^{0,0}(\tilde C)$ on the cone. 
\item[\textbf{3)}]\label{res:purity1} In Thm. \ref{thm6} we characterize the subclass of Hadamard states constructed in \textbf{2)}, which additionally are pure on the symplectic space $(\cH(\tilde C),\sigma_C)$ on the cone. It turns out that they can be parametrized by a single  operator in $\tilde\Psi^{-\infty,0}(\tilde{C})$.
\item[\textbf{4)}]\label{res:purity2} In Thm. \ref{thm66} we prove that if $\dim M\geq 4$, then the  pure states considered in \textbf{3)} induce pure states in the interior $M_0$ of the cone.
\end{enumerate}

In Subsect. \ref{ss:asymflat} we argue that Hypothesis \ref{hyp:main} covers  the case when $M_0$ is an asymptotically flat vacuum spacetime with future time infinity, after  a conformal transformation. Thus, our result \textbf{4)} solves an open question by Moretti \cite{Mo2} for $\dim M\geq4$.
\subsection*{Characteristic Cauchy problem} The proof of our main result \textbf{4)} relies on rather standard results on  the \emph{characteristic Cauchy problem} (also called \emph{Goursat problem} in the literature)  in appropriate Sobolev spaces.  We use 

Let $\Sigma$ be a Cauchy surface  for ($M, g$) in the future of $\{p\}$ and $\Sigma_0\defeq \Sigma\cap M_0$. We set 
\[
M_{1}\defeq I^{-}(\Sigma_{0}; M)\cap M_{0}, \ \ C_0\defeq ( J^-(\Sigma_0;M)\cap C)\cup \{p\},
\]
see Fig. 1. $M_{1}$ is relatively compact in $M$ with $\partial M_{1}= \Sigma_{0}\cup C_{0}$, 
 $\Sigma_0$ and $C_0$ are compact in $M$ with smooth boundary $\partial\Sigma_0=\partial C_0$. We denote by $H_0^1(\Sigma_0)$, $H_0^1(C_0)$ the respective restricted Sobolev spaces of order $1$, i.e. the space of distributions in $H^1(\Sigma_0)$, $H^1(C_0)$ that vanish on the boundary. \ \\
 \begin{center}
\includegraphics[scale=0.8]{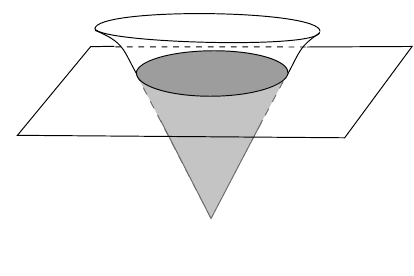}
\begin{picture}(0,0)\label{totop}
 \put(-90,15){$p$}
  \put(-80,75){$\Sigma_{0}$}
    \put(-85,55){$M_{1}$}
\put(-70,35){$C_{0}$}
  \put(-132,55){$\Sigma$}
\put(-95, 95){$C$}
\end{picture}\\
Fig. 1
\end{center}
If $f\in H_{0}^1(\Sigma_0)\oplus L^2(\Sigma_0)$  is a pair of Cauchy data, we denote by $e_{\Sigma_{0}}f$ its extension by $0$ to $\Sigma$ and by $u=U_{\Sigma_{0}}f$ the restriction to $M_{1}$ of the solution of the Cauchy problem
 \[
\begin{cases}
P u = 0\hbox{ in }M\\
\rho u = e_{\Sigma_{0}}f\hbox{ on }\Sigma,
\end{cases}
\]
where $\rho u= (u\tra{\Sigma}, \i^{-1}\p_{\nu}u\tra{\Sigma})$. By standard energy estimates  one obtains that $U_{\Sigma_{0}}: H_{0}^1(\Sigma_0)\oplus L^2(\Sigma_0)\to H^{1}(M_{1})$ is continuous.

 In Subsect. \ref{ss:pureint} we prove the following result.
\begin{theoreme}\label{thm:charcauchy} The map
\[
\begin{aligned}
T:H^1_0(\Sigma_0)\oplus L^2(\Sigma_0)\to &\, H^{1}_{0}(C_{0})\\
f\mapsto &\, (U_{\Sigma_{0}}f)\traa{C_{0}}
\end{aligned}
\]
is a homeomorphism. Moreover, if $\dim M\geq 4$ then $T(C_0^\infty(\Sigma_0)\oplus C_0^\infty(\Sigma_0) )$ is dense in $|D_s|^{-\12}L^2(\tilde C)$.
\end{theoreme}

{The first part of Thm. \ref{thm:charcauchy}  is equivalent to the existence and uniqueness of solutions in $M_{1}$  of the characteristic Cauchy problem:
\[
\begin{cases}
P u = 0,\hbox{ in }M_{1}, \\
u\tra{C_{0}} =  \varphi, \ \ \varphi\in H^{1}_{0}(C_{0}).
\end{cases}
\] 
The  proof proceeds by reduction to a case already considered by H\"ormander in \cite{Ho}, namely when the characteristic surface is the  graph of a Lipschitz function defined on a \emph{compact} domain.   
Beside  \cite{Ho} there is a considerable literature on the characteristic Cauchy problem for the Klein-Gordon equation,  for example \cite{BW,cagnac,dossa,nicolas2}, let us also mention related works on the Dirac equation \cite{nicolas,HN,joudioux}. The first part of Thm. \ref{thm:charcauchy} could actually also be deduced from \cite[Thm. 23]{BW}.}

The second part of Thm. \ref{thm:charcauchy} asserts that there is no loss of information on the level of purity of states when going from the cone $C$ to its interior $M_0$. The precise form of the statement comes from the fact that the one-particle Hilbert space associated to our Hadamard states, i.e. the completion of $\cH(\tilde C)$ for the inner product $(\cdot|(\lambda^++\lambda^-)\cdot)$, equals $|D_s|^{-\12}L^2(\tilde C)$. The validity of such result appears to be very delicate, it would be for instance problematic for $|D_s|^{-\alpha}L^2(\tilde C)$ with $\alpha<\12$ instead of $\alpha=\12$ and we do not know whether it holds for $d<3$. The generalization of Thm. \ref{thm:charcauchy} to other geometrical situations is thus an interesting open problem, particularly relevant for the quantum field theoretical bulk-to-boundary correspondence.

\subsection*{Plan of the paper} In Sect. \ref{sec6} we fix the geometric setup and outline the construction of null coordinates near the cone $C$. In Sect. \ref{sec10} we  briefly review the Klein-Gordon  field in $M_{0}$ and the definition of Hadamard states. Sect. \ref{bulktobound} is devoted to the so-called bulk-to-boundary correspondence, i.e. to the definition of a convenient symplectic space 
$(\cH(\tilde{C}), \sigma_{C})$ of functions on $C$, containing the traces on $C$ of space-compact solutions in $M_{0}$.

 In Sect. \ref{sec9}, we formulate the Hadamard condition on $C$, i.e. the natural microlocal condition on the two-point functions of a quasi-free state on $(\cH(\tilde{C}), \sigma_{C})$ which ensures that the induced state in $M_{0}$ is a Hadamard state.

Sect. \ref{sec20} is devoted to the pseudodifferential calculus on $\rr\times \bS$, more precisely to the `product-type' classes, associated to bi-homogeneous symbols. We also describe more general operator classes which are pseudodifferential only in the variables in  $\bS$. 

In Sect. \ref{sec21} we construct large classes of Hadamard states on the cone, whose covariances belong to the operator classes introduced in Sect. \ref{sec20}.  In Sect. \ref{sec22} we characterize {\em pure} Hadamard states, and show that they induce pure states in $M_{0}$. Finally in Sect. \ref{covar} we discuss the invariance of our classes of Hadamard states under change of null coordinates on $C$. Various technical results are collected in  Appendix \ref{secapp1}.
\section{Geometric setup}\label{sec6}\init
In this section we describe our geometrical setup and construct null coordinates near  the cone $C$.
\subsection{Future lightcone}\label{sec6.1}
We consider a globally hyperbolic spacetime $(M, g)$ of dimension $\dim M = d+1$.   \textcolor{Red}{If $K\subset M$, $I^{\pm}(K;M)$ resp. $J^{\pm}(K;M)$ denote the future/past  time-like resp. causal shadow  of $K$ in $M$, see e.g. \cite[Chap. 8]{W} or \cite[Sec. 1.3]{BGP} for more details. If  the spacetime $M$ is clear from the context these sets will simply be denoted by $I^{\pm}(K)$, $J^{\pm}(K)$.}

As outlined in the introduction, we fix a base point $p\in M$, and consider
\[
C= \p J^{+}(p)\backslash\{p\}, \ \ M_{0}=  I^{+}(p), 
\]
so that $C$ is the future lightcone from $p$, with tip removed, and $M_{0}$ is the interior of $C$. From \cite[Sect. 8.1]{W} we know that  $M_{0}$ is open, with 
\[
\overline{M_{0}}= J^{+}(p), \ \ \p M_{0}= \p J^{+}(p)= C\cup\{p\}.
\]
We assume Hypothesis \ref{hyp:main}, i.e. that there exists $f\in \cinf(M)$ such that:
\[
\begin{aligned}
(1) & \ C\subset f^{-1}(\{0\}), \ \nabla_{a}f\neq 0\hbox{ on }C,\ \nabla_{a}f(p)=0, \ \nabla_{a}\nabla_{b}f(p)= - 2 g_{ab}(p),\\[2mm]
(2) & \hbox{ the vector field }\nabla^{a}f \hbox{ is complete on }C.
\end{aligned}
\]
It follows that $C$ is a smooth hypersurface, although $\overline{C}$ is not smooth. Moreover since $C$ is  a null hypersurface, $\nabla^{a}f$ is tangent to $C$. 

\subsection{Causal structure}\label{sec6.1bb}
We now collect some  useful results on the causal structure of $M_0$ and $M$. 
\begin{lemma}\label{6.-1}
 Let $K\subset M_{0}$ be compact. Then:
 \beq\label{e6.01}J^{-}(K)\cap J^{+}(p)\hbox{ is compact},
\eeq
\begin{equation}
\label{e6.02}
J^{+}(K)\cap C= \emptyset.
\end{equation} 
\end{lemma}
\proof (\ref{e6.01}) follows from \cite[Lemma A.5.7]{BGP}. Moreover if $V\subset M_{0}$ is open with $K\subset V$, we have $J^{+}(K)\subset I^{+}(V)\subset M_{0}$. Since $\p J^{-}(p)= \p M_{0}$ and $M_{0}$ is open, this implies (\ref{e6.02}). \qed

\medskip

The following lemma is due to Moretti \cite[Thm. 4.1 (a)]{Mo1}. If $K\subset M_{0}$, the notation $J^\pm(K;M_0)$ or $J^\pm(K;M)$ is used in the place of $J^\pm(K)$ to specify which causal structure one refers to. 
 
\begin{lemma}\label{6.-2}
 The Lorentzian manifold $(M_{0}, g)$ is globally hyperbolic. Moreover
 \beq\label{e7.1}
 J^{+}(K; M_{0})= J^{+}(K; M), \ J^{-}(K; M_{0})= J^{-}(K; M)\cap M_{0}, \ \forall \ K\subset M_{0}.
\eeq
\end{lemma}

The next proposition is also due to Moretti \cite[Lemma 4.3]{Mo2}.
\begin{proposition}\label{6.-00}
 Let $K\subset M_{0}$ be compact. Then there exists a neighborhood $U_{1}$ of $p$ in $M$ such that no null geodesic starting from $K$ intersects $\overline{C}\cap U_{1}$.
\end{proposition}

\subsection{Asymptotically flat spacetimes}\label{ss:asymflat}
\def\tM{{M}}
\def\tg{{g}}

\newcommand{\scri}{{\mathscr I}}

In what follows we explain the relation between  Hypothesis \ref{hyp:main} and the geometrical assumptions met in the literature on Hadamard states \cite{Mo1,Mo2,DS,BDM}.

Let us consider two globally hyperbolic spacetimes $(M_0,g_0)$ and $(M,g)$, where $M_0$ is an embedded submanifold of $M$. One introduces the following set of assumptions. 

\begin{hypothesis}\label{hyp:aflat} Suppose the spacetime $(\tM, \tg)$ is such that:
\ben
\item there exists $\Omega\in C^{\infty}(\tM)$ with $\Omega>0$ on $M_0$ and $\tg\tra{M_0}= \Omega^{2}\traa{M_0}g_0$,
\item there exists $i^{-}\in \tM$ such that $J^{+}(i^{-}; \tM)$ is closed and 
\[
M_0= J^{+}(i^{-}; \tM)\backslash \p J^{+}(i^{-}; \tM),
\]
 \item $g_0$ solves the vacuum Einstein equations at least in a neighborhood of $\scri^-$
 \[
 \scri^{-}\defeq\p J^{+}(i^{-}; \tM)\backslash \{i^{-}\},
 \]
 \item $\Omega=0$ and $d\Omega\neq 0$ on $\scri^{-}$, $d\Omega(i^{-})=0$, $\nabla_{a}\nabla_{b} \Omega(i^{-})= - 2 \tg_{ab}(i^{-})$,
 \item if $n^{a}\defeq\tg^{ab}\nabla_{b}\Omega$, then there exists $\omega\in \cinf(\tM)$, with $\omega>0$ on $M_0\cup \scri^{-}$ and 
 
{\upshape (a)}  $\nabla_{a}(\omega^{4}n^{a})=0$ on $\scri^{-}$, 

{\upshape (b)} the vector field $\omega^{-1}n$ is complete on $\scri^{-}$,
\een
\end{hypothesis}

Above, the symbols $\nabla_a$ refer to the metric $g$.

One says that $(M_0,g_0)$ is an \emph{asymptotically flat vacuum spacetime with past time infinity $i^-$} if there exists a spacetime $(M,g)$ such that $M_0$ is an embedded submanifold of $M$ and Hypothesis \ref{hyp:aflat} is satisfied\footnote{Note that we consider here only globally hyperbolic spacetimes, cf. \cite[App. A]{Mo2} for a more general definition.}.

\begin{lemma}\label{lem:flat} Suppose $(M_0,g_0)$ is an asymptotically flat vacuum spacetime with past time  infinity $i^-$ and let $(M,g)$ satisfy Hypothesis \ref{hyp:aflat}. Then Hypothesis \ref{hyp:main} is satisfied for $p\defeq i^-$ and $f=\omega\Omega$. 
\end{lemma}
Note that actually only conditions (1), (2), (4) and (5b) in Hypothesis \ref{hyp:aflat} are needed in Lemma \ref{lem:flat}.

{\Red In the present paper we construct Hadamard states for the Klein-Gordon operator $P=-\Box_g +r(x)$ in $(M_0,g\traaa{M_0}\,)$ for any smooth, real valued $r$. In the special case of the conformal wave operator $P=-\Box_g + \frac{n-2}{4(n-1)}R$ (with $R$ the scalar curvature) this yields however also Hadamard states on $(M_0,g_0)$ since the two metrics are conformally related, cf. Appendix \ref{ss:confo}.}


\subsection{Null coordinates near $C$}\label{sec6.1b}
For later use it is convenient  to introduce null coordinates near $C$.    The construction seems to be well-known, we sketch it for the reader's convenience.  Note however the estimates in Lemma \ref{7.1}, which will be useful later on.

We first choose normal coordinates $(y^{0}, \overline{y})$ at $p$  such that on  a neighborhood  of $p$,  $C= \{ (y^{0})^{2}- |\overline{y}|^{2}=0, \ y^{0}>0\}$.

Set 
\beq\label{e6.fourest}
v\defeq  y^{0}+ | \overline{y}|, \ w\defeq  y^{0}- |\overline{y}|, \ \psi\defeq  \frac{\overline{y}}{|\overline{y}|}\in \bS,
\eeq
so that  on a neighborhood of $p$ one has $C= \{ w=0, v>0\}$.  Abusing notation slightly, we denote by $\psi^{1}, \dots, \psi^{d-1}$ coordinates on  $\bS$,  and use the same letter for their pullback to local coordinates on $M$ near $p$. 
We set  \beq\label{def-de-surf}
S\defeq  \{w=0, \ v= \epsilon_{0}\}
\eeq where  $\epsilon_{0}>0$  will be chosen small enough. Note that  $S\subset C$ is diffeomorphic to $\bS$.

\begin{lemma}\label{7.1}
\ben
\item There exists a unique solution $s\in \cinf(C)$ of:
\[
\begin{cases}
(\nabla^{a}f \nabla_{a}s)\tra{C}=-1,\\
s\tra{S}=0.
\end{cases}
\]
\item There exists unique solutions $\theta^{j}\in\cinf(C)$, $1\leq j\leq d-1$ of:
\[
\begin{cases}
(\nabla^{a}f \nabla_{a}\theta^{j})\tra{C}=0,\\
\theta^{j}\tra{S}=\psi^{j}.
\end{cases}
\]
\item Moreover there exists $0<\epsilon_{0}<\epsilon_{1}$ and $k, \tilde{\theta}^{j}\in \cinf(]-\epsilon_{1}, \epsilon_{1}[\times\bS)$ such that 
 \[
s(v, \psi)= \12\ln(v)+ k(v, \psi), \ \theta^{j}(v, \psi)= \tilde{\theta}^{j}(v, \psi), \hbox{ on }]0, \epsilon_{0}[\times\bS.
\]
\een
\end{lemma}
\proof  The proof is given in Appendix \ref{fred-crook}. \qed


\medskip

It remains to extend $s, \theta^{j}$ to smooth functions on a neighborhood of $C$.

We  argue as in \cite[Sect. 11.1]{W}: for $s_{0}\in \rr$, the submanifold $S_{s_{0}}= \{s= s_{0}\}\subset C$ is spacelike, of codimension $2$ in $M$. At a given point  of $S_{s_{0}}$ the orthogonal to its tangent space is two dimensional, timelike, and hence contain two null lines. One of them is generated by $\nabla^{a}f$, the other is transverse to $C$. 
We extend  $(s, \theta)$ to  a neighborhood of $C$ by imposing that $(s, \theta)$ are constant along the  above family of null geodesics, transverse to $C$.
\begin{lemma}\label{taz}
 The functions $(f, s, \theta)$ constructed above are a system of local coordinates near $C$ with $C\subset \{f=0\}$ and
\begin{equation}
\label{e6.03}
g\tra{C}= -2df ds+ h_{ij}(s, \theta)d\theta^{i}d\theta^{j},
\end{equation}
where $h_{ij}(s, \theta)d\theta^{i}d\theta^{j}$ is a smooth, $s-$dependent Riemannian metric on $\mathbb{S}^{d-1}$.
\end{lemma}
\proof The proof will be given in  Appendix \ref{proof-of-lemid}. \qed
\subsection{Estimates on traces}\label{sec6.1.c}
In this subsection we derive estimates, in the coordinates $(s, \theta)$  on $C$ constructed above, for the restriction to $C$ of a smooth, space compact function in $M$. These estimates will be applied later to traces on $C$ of solutions of the Klein-Gordon equation in $M_{0}$.

We recall that $\cinf_{\rm sc}(M)$ denotes the space of smooth {\em space compact }functions i.e. the space of $\phi\in\cinf(M)$ such that $\supp \phi\subset J^{+}(K)\cup J^{-}(K)$ for some compact $K\subset M$.

We will slightly abuse notation by writing $\phi(x^{0}, \dots, x^{d})$ for the function $\phi$ expressed in some coordinate system $(x^{0}, \dots, x^{d})$ near $p$. We will similarly write for example $\phi(v, \psi), \phi(s, \theta)$ for $\phi\in\cinf(C)$.

By Lemma \ref{6.-1} we see that $\supp \phi\cap \overline{C}$ is compact in $\overline{C}$ if $\phi\in  \cinf_{\rm sc}(M)$. This means that it suffices to control the derivatives in $(s, \theta)$ of $\phi\tra{C}(s, \theta)$ near $s=-\infty$, i.e. of $\phi\tra{C}(v, \psi)$ near $v=0$.
Clearly the only task is to control what happens near $p$, i.e. when $s\to -\infty$.  
We first derive estimates in the coordinates $(v, \psi)$ introduced in (\ref{e6.fourest}), in a neighborhood of $v=0$. If $\phi\in \cinf_{\rm sc}(M)$ we denote by $\phi(y^{0}, \overline{y})$  the function $\phi$ expressed in normal coordinates at $p$, which is defined on a neighborhood of $0$. We set then
\[
\hat{\phi}(v, \psi)= \phi(\textstyle\12 v, \textstyle\12 v \psi)\in \cinf(]-\epsilon_{1}, \epsilon_{1}[\times \bS), \hbox{ for some }\epsilon_{1}>0,
\]
so that
\[
\phi\tra{C}= \hat{\phi}\tra{\{v>0\}}
\]

We denote by $S^{0}$ the space of functions $u(v, \psi)\in \cinf(]-\epsilon_{1}, \epsilon_{1}[\times \bS)$ which are bounded with all derivatives.

\begin{lemma}\label{7.2}
 \ben
 \item if $\phi\in \cinf_{{\rm sc}}(M)$ then $ \hat{\phi}(v, \psi)$ belongs to $S^{0}$.
 \item  Let $|h|= \det[h_{ij}]$. Then $|h|(v, \psi)= v^{2(d-1)} r_{0}(v, \psi)$ for $r_{0}, r_{0}^{-1}\in S^{0}$.
 \een
\end{lemma}

\proof 
  Considering the map $\chi: \bS\ni \psi\mapsto \psi\in  \rr^{d}$ and denoting still by $\psi$ some coordinates on $\bS$ we have:
\[
\p_{v}\tilde{\phi}= \12(\p_{y^{0}}\phi- \psi\cdot \p_{\overline{y}}\phi), \ \p_{\psi^{i}}\tilde{\phi}= \12 v \p_{\psi^{i}}\chi^{j}\p_{\overline{y}^{j}}\phi.
\]
From this we obtain (1). To prove (2) we need to express $h_{ij}= \langle\p_{\theta^{i}}| g \p_{\theta^{j}}\rangle$ on $C$. An easy computation using the estimates in Lemma \ref{7.1} shows that on $C$ we have:
\[
\p_{\theta^{i}}= a_{i}^{j}(v, \psi)\p_{\psi^{j}}+ v r_{0}(v, \psi)\p_{v},
\]
where $a^{j}_{i}, r_{0}\in S^{0}$ and $[a^{ij}](v, \psi)$ invertible.
Plugging this into (\ref{e7.2}), we obtain
\[
[h_{ij}](v, \psi)=  v^{2}\left(^{t}[a_{i}^{j}](v, \psi)[m_{ij}](\psi)[a_{i}^{j}](v, \psi)+ v [b_{ij}](v, \psi)\right),
\]
where $b_{ij}\in S^{0}$. This implies (2). \qed

\medskip


%
%
%
We will also need later the following lemma. We denote by $m_{ij}(\theta)d\theta^{i}d\theta^{j}$ the standard Riemannian metric on $\bS$ and set:
\begin{equation}
\label{e7.4b}
\beta(s, \theta)\defeq  |m|^{\frac{1}{4}}(\theta)|h|^{-\frac{1}{4}}(s, \theta),
\end{equation}
\begin{lemma}\label{7.3}
 Let 
 \[
\tilde{\phi}(s, \theta)\defeq  \beta^{-1}(s, \theta)\phi\traaa{C}(s, \theta), \ \phi\in \cinf_{\rm sc}(M),
\]
Then for all $s_{1}\in \rr$ one has:
\[
 \p_{s}^{\alpha}\p_{\theta}^{\beta}\tilde{\phi}\in O(\e^{s(d-1)}), \ s\in ]-\infty,  s_{1}], \ \forall \ \alpha, \beta.
\]
\end{lemma}
\proof
We note that $\beta^{-1}= v^{(d-1)/2}r_{0}(v, \psi)$, for $r_{0}, r_{0}^{-1}\in S^{0}$.  From this and Lemma \ref{7.2}  it follows that if $\phi\in \coinf(M)$, then $\tilde{\phi}(v, \psi)\in  v^{(d-1)/2}S^{0}$. It remains to estimate the derivatives of $\tilde{\phi}$ w.r.t. $s$ and $\theta$.
By a standard computation we obtain for $u\in \cinf(]-\epsilon_{1}, \epsilon_{1}[\times \bS)$:
\[
\begin{aligned}
\p_{\theta^{i}}u &= a^{j}_{i}(v, \psi)\p_{\psi^{j}}u+ v r_{i}(v, \psi)\p_{v}u, \\[2mm]
\p_{s}u &= v(1+ v r_{0}(v, \psi)) \p_{v}u+ v b^{j}(v, \psi)\p_{\psi^{j}}u,
\end{aligned}
\]
for $r_{0}, r_{i}, b^{j}, a^{j}_{i}\in S^{0}$, and $[a^{j}_{i}]$ invertible. From this point on the lemma is a routine computation. \qed

\section{Klein-Gordon fields inside the future lightcone}\label{sec10}\init
\subsection{Klein-Gordon equation in $M_{0}$}
 We  fix a smooth real function $r\in \cinf(M)$ and consider the {\em Klein-Gordon operator} on $(M, g)$:
 \[
P(x, D_{x})= -\nabla^{a}\nabla_{a}+ r(x),\hbox{ acting on }\cinf(M).
\]
We denote by $E_{\pm}\in \cD'(M\times M)$ the retarded/advanced Green's functions for $P$, by $E= E_{+}- E_{-}\in \cD'(M\times M)$ the Pauli-Jordan commutator function,  and by $\Sol(P)$ the space of smooth, complex valued, space-compact solutions of 
\[
P(x, D_{x})\phi=0\hbox{ in }M.
\]
Recall  that we have set in Subsect. \ref{sec6.1}:
 \[
M_{0}\defeq  I^{+}(p),
\]
 and by Lemma \ref{6.-2} we know that $(M_{0}, g)$ is globally hyperbolic.

We denote by $P_{0}= -\nabla^{a}\nabla_{a}+ r(x)$ the restriction of $P$ to $M_{0}$,  by $E_{0}\in \cD'(M_{0}\times M_{0})$ the Pauli-Jordan function for $P_{0}$, and by $\Sol(P_{0})$ the space of smooth, complex valued, space-compact solutions of 
\[
P_{0}(x, D_{x})\phi_{0}=0\hbox{ in }M_{0}.
\]
By  the global hyperbolicity of $(M_{0}, g)$ we know that $\Sol(P_{0})= E_{0}\cD(M_{0})$. 
From (\ref{e7.1})  and the uniqueness of $E_{0\pm }$ we obtain that $E_{0\pm}= E_{\pm}\,\traa{ M_{0}\times M_{0}}$, hence 
\[
E_{0}= E\tra{M_{0}\times M_{0}}.
\] 
It follows that any $\phi_{0}\in \Sol(P_{0})$ uniquely extends to  $\phi\in \Sol(P)$, in fact 
\beq\label{e9.1}
\phi_{0}= E_{0} f_{0}, \ f_{0}\in \cD(M_{0})\Rightarrow \phi_{0}= Ef_{0}\traa{ M_{0}}.\eeq
As usual we  equip $\Sol(P_{0})$ with the symplectic form
\beq\label{e7.5}
 \overline{\phi}_{1}\sigma_{0} \phi_{2}\defeq  \int_{\Sigma_{0}} \overline{\nabla_{a}\phi_{1}} \phi_{2}- \overline{\phi}_{1}\nabla_{a}\phi_{2}n^{a} d\sigma_{h},
\eeq
where $\Sigma_{0}\subset M_{0}$ is a Cauchy hypersurface for $(M_{0}, g)$ (see Subsect. \ref{secapp1.1} for notation).
It is well known that
\[
E_{0}: (\coinf(M_{0})/P_{0}\coinf(M_{0}), E_{0})\to (\Sol(P_{0}), \sigma_{0})
\]
is a symplectomorphism.
\subsection{Hadamard states in $M_{0}$}
We first briefly recall some standard facts, and refer for example to \cite[Sect. 2]{GW} for details and notation.

\subsubsection{Covariances of a quasi-free state}\label{ss:qfree}
If  $(\cY, \sigma)$ is a complex symplectic space,  the {\em complex covariances} $\Lambda^{\pm}\in L_{\rm h}(\cY, \cY^{*})$ of a (gauge invariant) quasi-free state $\omega$ on $\CCR(\cY, \sigma)$ (the polynomial CCR $^*$-algebra of $(\cY,\sigma)$) are defined by:
\[
\omega(\psi(y_{1})\psi^{*}(y_{2}))\eqdef  (y_{1}| \Lambda^{+}y_{2}), \ \ \omega(\psi^{*}(y_{2})\psi(y_{1}))\eqdef  (y_{1}| \Lambda^{-}y_{2}), \ \ y_{1}, y_{2}\in \cY.
\]
From the CCR we obtain that $\Lambda^{+}- \Lambda^{-}= \i \sigma\eqdef q$, and the necessary and sufficient condition for $\Lambda^{\pm}$ to be the complex covariances of a (gauge invariant) quasi-free state is  that $\Lambda^{\pm}\geq 0$.

If $(\cY, \sigma)= (\coinf(M_{0})/P\coinf(M_{0}), E_{0})$, the complex covariances of a state $\omega$ are induced from {\em two-point functions}, still denoted by $\Lambda^{\pm}$ such that
\[
 \Lambda^{\pm}\in \cD'(M_{0}\times M_{0}),\quad P \Lambda^{\pm}= \Lambda^{\pm}P=0,
\]
where we identify operators on $\coinf(M_{0})$ with sesquilinear forms using the scalar product 
\[
(u|v)\defeq  \int_{M_{0}}\bar{u}v \,d\mu_{g}, \ u, v\in \coinf(M_{0}).
\]
\subsubsection{Hadamard condition}\label{hadcond}
We now recall the {\em Hadamard condition} for quasi-free states.
We denote by $T^{*}M$ the cotangent bundle of $M$ and $Z= \{(x, 0)\}\subset T^{*}M$ the zero section. 
 The {\em principal symbol} of $P$ is
$p(x, \xi)=\xi_{a}g^{ab}(x)\xi_{b}$, the set 
 \[
\cN\defeq  \{(x, \xi)\in \coo{M}: p(x, \xi)=0\}
\]
 is called the  {\em characteristic manifold} of $p$.

 The Hamilton vector field of $p$ will be denoted by $H_{p}$, whose integral curves inside $\cN$ are called {\em bicharacteristics}. 
 
 We  will  use the notation $X=(x,\xi)$ for points in $\coo{M}$ and  write $X_1\sim X_2$ if $X_1=(x_1,\xi_1)$ and $X_2=(x_2,\xi_2)$ are in $\cN$ and $X_1$ and $X_2$ lie on the same bicharacteristic  of $p$.

Let us  fix a time orientation and denote  by $V_{x\pm}\subset T_x M$ for $x\in M$, the open future/past light cones and $V_{x\pm}^*$ the dual cones
\[
V^{*\pm}_{x}\defeq \{ \xi\in T^*_{x}M : \ \xi\cdot v >0, \ \forall \,v\in V_{x\pm}, \ v\neq 0 \}.
\]
The set $\cN$ has two connected components invariant under the Hamiltonian flow of $p$, namely:
\[
\cN^{\pm}\defeq \{ X\in \cN : \ \xi\in V^{*\pm}_{x}\}.
\]
\begin{definition}
A quasi-free state $\omega$ on $\CCR(\coinf(M_{0})/P\coinf(M_{0}), E_{0})$ with two-point functions $\Lambda^{\pm}$ satisfies the {\em microlocal spectrum condition} if:
 \beq\label{musc}\tag{$\mu$sc}
\WF(\Lambda^{\pm})'\subset \cN^{\pm}\times \cN^{\pm}.
\eeq
Quasi-free states satisfying (\ref{musc}) are called {\em Hadamard states}.
\end{definition}

{\Red This form of the Hadamard condition was shown in \cite{SV} to be equivalent to older definitions \cite{radzikowski}, we refer the reader to \cite{sanders,wrothesis} for a discussion on equivalent formulations of the microlocal spectrum condition.}

\section{Bulk-to-boundary correspondence}\label{bulktobound}
\subsection{Boundary symplectic space}\label{sec10.2}
\def\tC{\tilde{C}}
We equip $C$ with the coordinates $(s, \theta)$ constructed in Subsect. \ref{sec6.1b} 	and hence identify $C$ with \beq\label{def-de-tC}
\tC\defeq  \rr\times\bS.
\eeq
We denote by $H^{k}(\tC)$, $k\in \nn$ the Sobolev space 
\[
H^{k}(\tC)\defeq  \big\{g\in \cD'(\rr\times\bS): \int |\p^{\alpha}_{s}\p^{\beta}_{\theta} g|^{2}|m|^{\12}d s d \theta <\infty, \ \alpha +|\beta|\leq k  \big\},
\]
and extend the definition of $H^{k}(\tC)$ to $k\in \rr$ in the usual way.  The space $H^{0}(\tC)$ will be denoted simply by $L^{2}(\tC)$. We set also:
\[
\cH(\tC)\defeq  \bigcap_{k\in \rr}H^{k}(\tC), \quad \cH'(\tC)\defeq  \bigcup_{k\in \rr}H^{k}(\tC),
\]
equipped with their canonical topologies.

We set
\begin{equation}
\label{e10.1}
 \overline{g}_{1}\sigma_{C}  g_{2}\defeq  \int_{\rr\times\bS} ( \p_{s}\overline{g}_{1}g_{2}-\overline{g}_{1} \p_{s}g_{2})|m|^{\12}(\theta) ds d\theta, \ g_{1}, g_{2}\in \cH(\tC).
\end{equation}

Introducing    the {\em charge} $q\defeq  \i \sigma_{C}$ we have:
\[
 \overline{g}_{1}q g_{2}= 2(g_{1}| D_{s} g_{2})_{L^{2}(\tC)}, \ g_{1}, g_{2}\in  \cH(\tC),
\]
where $D_{s}= \i^{-1}\p_{s}$ is  selfadjoint on $L^{2}(\tC)$ on its natural domain. Clearly   $(\cH(\tC), \sigma_{C})$ is a complex symplectic space.
\subsection{Bulk-to-boundary correspondence}\label{butobound}

\begin{definition}\label{10.1}
Let $\beta\in \cinf(\tC)$ be defined in (\ref{e7.4b}). We set
\[
\begin{array}{rl}
\rho:& \Sol(P_{0})\to \cinf(\rr\times\bS)\\[2mm]
& \phi\mapsto \beta^{-1}(s, \theta) \phi\traaa{C}(s, \theta).
\end{array}
\] 
\end{definition} 
\begin{proposition}\label{10.2}
 \ben
 \item $\rho$ maps $\Sol(P_{0})$ into $\cH(\tC)$;
 \item $\rho: (\Sol(P_{0}), \sigma)\to (\cH(\tC), \sigma_{C})$ is a monomorphism, i.e.:
 \[
\overline{\rho\phi}_{1}\sigma_{C}\rho\phi_{2}= \overline{\phi}_{1}\sigma \phi_{2}, \ \forall \, \phi_{1}, \phi_{2}\in \Sol(P_{0}).
\]
 \een
\end{proposition}
\proof  Let $\phi_{0}, \phi$ as in (\ref{e9.1}). By Lemma \ref{6.-1} and the support properties of $E$, we see that $\supp \phi\cap \overline{C}$ is compact in $M$. Therefore  the restriction of $\phi$ to $C$ equals  the restriction of a smooth compactly supported function to $C$.  By Lemma \ref{7.3}   and the fact that $\rho\phi_{0}$ is supported in $]-\infty, s_{1}]\times\bS$ for some $s_{1}$, we obtain that $\rho\phi_{0}\in \cH(\tC)$, which proves (1).
 
 We now prove (2). Let $\phi_{i, 0}\in \Sol(P_{0})$, $i=1,2$ which are restrictions to $M_{0}$ of  $\phi_{i}\in \Sol(P)$. We  fix a Cauchy surface $\Sigma_{0}$  for $(M_{0}, g)$ such that $\supp \phi_{i,0}\cap \Sigma_{0}\subset K\Subset M_{0}$. We can find a Cauchy surface $\Sigma$ for $(M, g)$ such that $\Sigma\cap K= \Sigma_{0}\cap K$.
Denoting by
\[
J_{a}(\phi_{1}, \phi_{2})\defeq  \overline{\phi}_{1}\nabla_{a}\phi_{2}- \overline{\nabla_{a}\phi_{1}} \phi_{2},
\]
the conserved current, we have:
\[
\overline{\phi}_{1, 0}\sigma_{0} \phi_{2, 0}= \overline{\phi}_{1}\sigma \phi_{2}, 
\]
where
\[
 \overline{\phi}_{1}\sigma \phi_{2}=- \int_{\Sigma} J_{a}(\phi_{1}, \phi_{2})n^{a} d\sigma_{h},
\]
is the symplectic form on $\Sol(P)$. We now apply Stokes formula  in the form (\ref{e.app3}) to the domain $U\subset M$ bounded by $\Sigma\cap K$, $\overline{C}$ and $\p J^{+}(\Sigma\cap K)$, using that $\nabla_{a}J^{a}(\phi_{1}, \phi_{2})=0$. The boundary term on $\Sigma\cap K$ yields $ - \overline{\phi}_{1}\sigma \phi_{2}$, the boundary term on $\p J^{+}(\Sigma\cap K)$ vanishes.  To express the boundary term on $\overline{C}$, we use the coordinates $(f, s, \theta)$ constructed in Subsect. \ref{sec6.1b}. We  formally obtain the quantity:
\[
 \overline{g}_{1}\hat \sigma g_{2}=\int_{\rr\times\bS}( \p_{s}\overline{g}_{1}g_{2}-\overline{g}_{1} \p_{s}g_{2})|h|^{\12}(s, \theta) ds d\theta
\]
for $g_{i}= (\phi_{i}) \traa{C}$.   
This equals $\overline{\rho\phi}_{1}\sigma_{C}\rho\phi_{2}$ by an easy computation.

To justify the use of Stokes formula, we need to take care of the fact that  $\overline{C}$ is not  smooth at $p$. This can be done as follows: for $0<\epsilon\ll 1$ we denote by $U_{\epsilon}$ some $\epsilon-$neighborhood of $p$. We  replace $\overline{C}$ by   a smooth hypersurface $C_{\epsilon}$, obtained by  smoothly gluing $C\backslash U_{\epsilon}$ to a 
piece of  a Cauchy surface  $\Sigma'_{\epsilon}$ passing through $U_{\epsilon}$. The contribution of the integral on $\Sigma_{\epsilon}$ is written using (\ref{e3.2}), and converges to $0$ when $ \epsilon\to 0$, using that $\phi_{i}$ are smooth functions. The contribution of the integral on $C\backslash U_{\epsilon}$ converges  to $\overline{\rho\phi}_{1}\sigma_{C}\rho\phi_{2}$, using that $\rho\phi_{i}\in \cH(\tC)$. This completes the proof of the proposition. \qed

\subsection{Pullback of states from the boundary}
Since 
\[
\rho:  (\Sol(P_{0}), \sigma_{0})\to (\cH(\tC), \sigma_{C})
\]
 is a monomorphism, we can pullback a quasi-free state  $\omega_{C}$ on $\CCR(\cH(\tC), \sigma_{C})$ to a  quasi-free state $\omega_{0}$ on $\CCR(\coinf(M_{0})/P_{0}\coinf(M_{0}), E_{0})$ by setting:
\begin{equation}
\label{pullpull}
\omega_{0}(\psi(u_{1})\psi^{*}(u_{2}))\defeq  \omega_{C}\,(\psi(\rho\circ E_{0} u_{1})\psi^{*}(\rho\circ E_{0}u_{2})), \ u_{1}, u_{2}\in\coinf(M_{0}).
\end{equation}
If $\lambda^{\pm}\in L_{\h}(\cH(\tC), \cH(\tC)^{*})$ are the complex covariances of $\omega_{C}$, then the complex covariances of $\omega_{0}$ are (formally) given by:
\beq\label{defdeLam}
\Lambda^{\pm}\defeq  (\rho\circ E_{0})^{*}\circ \lambda^{\pm} \circ (\rho\circ E_{0}).
\eeq
\section{Hadamard condition on the cone}\label{sec9}\init
In this section we formulate the natural boundary version of the bulk Hadamard condition (\ref{musc}).
\subsection{Preparations}\label{sec8.1}
We recall that $p(x, \xi)$ denotes the principal symbol of the Klein-Gordon operator $P$ (or $P_{0}$).
 
 Let $C\subset M$ be the forward lightcone introduced in Subsect. \ref{sec6.1}. We denote by $N^*C\subset \coo{M}$ the {\em conormal bundle} to $C$, i.e. 
\[
N^*C\defeq \{(x, \xi)\in \coo{M}: \ x\in C ,\  \xi=0\hbox{ on }T_{x}C\}.
\] 
The fact that $C$ is characteristic is equivalent to 
\beq\label{e8.2}
N^*C\subset \cN,
\eeq  
where $\cN$ is the characteristic manifold of $p$.  
Since $N^*C$ is Lagrangian,   it is well known that (\ref{e8.2}) implies that $N^*C$ is invariant under the flow of $H_{p}$. The projections on $M$ of bicharacteristics  starting from $N^*C$ are (modulo reparametrization) {\em characteristic curves}, i.e. integral curves of the vector field $v^{a}=\nabla^{a}f$, if $f\in \cinf(M)$ is some  defining function of $C$, i.e. $f=0$, $df\neq 0$ on $C$.

We will  use the coordinates $(f, s, \theta)$ introduced in Subsect. \ref{sec6.1b}, which, for ease of notation,   will be denoted by $x= (r, s, y)\in \rr\times \rr\times\bS$. The dual coordinates are denoted $ \xi= (\varrho, \sigma, \eta)$, elements of $T^{*}M$ will sometimes be denoted by $X= (x, \xi)$ and elements of $T^{*}C$ will be denoted by $Y= ((s, y), (\sigma, \eta))$.

In  the above coordinates, we have
\[
C=\{r=0\}, \ N^*C=\{r=0, \sigma=\eta=0\},
\]
and from (\ref{e6.03}) we obtain that:
\beq\label{e8.7b}
p(x, \xi)\tra{C}=  - 2\varrho \sigma+ h(s, y, \eta),
\eeq
where we set $h(s, y, \eta)= h^{ij}(0, s, y)\eta_{i}\eta_{j}$. Note that $h(s, y, \eta)$ is {\em elliptic}, i.e. $h(s, y, \eta)\geq c_{0}|\eta|^{2}$, for $c_{0}>0$, locally in $(s, y)$, since  $h_{ij}dy^{i}dy^{j}$ is Riemannian.

For later use let us extend the notation $X_{1}\sim X_{2}$ introduced in \ref{hadcond}. For   $Y= (s,y, \sigma, \eta)\in T^{*}C$ , $X= (x, \xi)\in T^{*}M$, we will  write $Y\sim X$ if
\beq\label{e8.6b}
\sigma\neq 0,\ 
((0, s, y), ((2\sigma)^{-1}h(s, y,\eta ),\sigma,  \eta)) \sim X.
\eeq
Recall also that the positive/negative energy components $\cN^{\pm}$ of $\cN$ were defined in \ref{hadcond}.\begin{lemma}\label{8.2b}
 Let $Y_{1}= (s_{1},y_{1}, \sigma_{1}, \eta_{1})\in T^{*}C$, $X_{2}= (x_{2}, \xi_{2})\in T^{*}M$ with $x_{2}\not\in C$. Then:
 \ben
 \item  there exists $\varrho_{1}\in \rr$ such that 
 \[
 X_{1}\defeq ((0, s_{1}, y_{1}), (\varrho_{1}, \sigma_{1}, \eta_{1}))\sim (x_{2}, \xi_{2})\eqdef X_{2}
 \] iff $\sigma_{1}\neq 0$ and then $\varrho_{1}= (2\sigma_{1})^{-1}h(s_{1}, y_{1},\eta_{1} )$ and $Y_{1}\sim X_{2}$.
 \item if $Y_{1}\sim X_{2}$ then $ X_{2}\in \cN^{\pm}$ iff $\pm \sigma_{1}>0$.
 \een
\end{lemma}
\proof 
Let $X_{1}= ((0, s_{1}, y_{1}), (\varrho_{1}, \sigma_{1}, \eta_{1}))\in \cN$. By (\ref{e8.7b}) we have
\[
 - 2\varrho_{1}\sigma_{1}+ h(s_{1}, y_{1}, \eta_{1})=0.
\] 
If $\sigma_{1}=0$ then $h(s_{1}, y_{1}, \eta_{1})=0$ hence $\eta_{1}=0$ by ellipticity of $h$. Therefore $\sigma_{1}=0$ implies $ X_{1}\in N^*C$.
 Since $X_{2}\sim X_{1}$ and $N^*C$ is invariant under the flow of $H_{p}$,  we have also $X_{2}\in N^*C$  which contradicts the hypothesis that $x_{2}\not\in C$.
 Therefore necessarily $\sigma_{1}\neq 0$ and hence $\varrho_{1}= (2\sigma_{1})^{-1} h(s_{1}, y_{1}, \eta_{1})$ and $Y_{1}\sim X_{2}$. This proves (1).

 To prove (2) we have to show that 
 \beq\label{e8.6c}
 \pm \sigma_{1}>0\Leftrightarrow ((0, s_{1}, y_{1}), ((2\sigma_{1})^{-1} h(s_{1}, y_{1}, \eta_{1}), \sigma_{1}, \eta_{1}))\in \cN^{\pm}.
\eeq
Let  us fix $(y_{1}, \eta_{1})\in T^{*}\bS$, $\sigma_{1}\in \rr$.   Since $\cN^{\pm}$ are the two connected components of $\cN$, it suffices by connexity to prove (\ref{e8.6c}) for $s_{1}$ in a neighborhood of $-\infty$, i.e. in a neighborhood of $p$ in $M$.  Recall that we introduced Gaussian normal coordinates $(y^{0}, \overline{y})$ near $p$ with  $\p_{y^{0}}$ future oriented. Let $\alpha$ be the one form $ (2\sigma_{1})^{-1}h(s_{1}, y_{1}, \eta_{1})dr+ \sigma_{1}ds+ \eta_{1}dy$. Then 
\[
((0, s_{1}, y_{1}), ((2\sigma_{1})^{-1} h(s_{1}, y_{1}, \eta_{1}), \sigma_{1}, \eta_{1}))\in \cN^{\pm}\Leftrightarrow \mp \langle \alpha| g^{-1} dy^{0}\rangle >0.
\]
Since it suffices to check the sign of $\langle \alpha| g^{-1} dy^{0}\rangle$ near $p$, we can, by a simple approximation argument (see e.g. (\ref{e7.2}))
replace $g$ by the flat metric at $p$. We have then (see Lemma \ref{7.1} and recall that $s=u, r=f$):
\[
y^{0}= v+w,  \ v= \e^{s}, \ w= \e^{-s }r,
\]
hence
\[
\mp\langle \alpha| g^{-1} dy^{0}\rangle=  \pm2(\e^{-s_{1}}\sigma_{1}+ \e^{ s_{1}}(2 \sigma_{1})^{-1}h(s_{1}, y_{1}, \eta_{1})),
\]
has the same sign as $\pm\sigma_{1}$, which proves (\ref{e8.6c}). \qed

\medskip

Recall that $E\in \cD'(M\times M)$ is the Pauli-Jordan commutator function for $P$  and  $\rho:\cD(M)\ni u\mapsto u\traa{C}\in C^\infty(\tC)$ is (modulo a smooth, non-zero multiplicative factor) the operator of restriction to $C$, defined in Def. \ref{10.1}.

{\Red Let us recall some notation: identifying $\co{(M_{1}\times M_{2})}$ with $\co{M_{1}}\times \co{M_{2}}$ one  denotes by $(\co{M_{1}}\times \co{M_{2}})\backslash Z$ the image of $\coo{(M_{1}\times M_{2})}$ under this identification.   If $\Gamma\subset (T^{*}M_{1}\times T^{*}M_{2})\backslash Z $ one sets: 
\beq\label{eq:newnotation}
\begin{aligned}
_{M_{1}}\!\Gamma&\defeq\{(x_{1}, \xi_{1}) : \ \exists \ x_{2}\hbox{ s.t. } (x_{1}, \xi_{1}, x_{2},0)\in \Gamma\} \subset \coo{M_{1}}_{1},\\
\Gamma\!_{M_{2}}&\defeq\{(x_{2}, \xi_{2}) : \ \exists \ x_{1}\hbox{ s.t. } (x_{1}, 0, x_{2},\xi_{2})\in \Gamma\}\subset\coo{M_{2}}_{2},
\end{aligned}
\eeq	
where $Z_i$ is the zero section of $T^{*}M_{i}$.}
\begin{proposition}\label{8.2}
 Let $\chi\in \coinf(M)$ with $\supp\chi\subset M\backslash C$ and $\psi\in \coinf(\tC)$. Then:
 \ben
 \item
 $\quad\WF(\psi \rho\circ E \chi)'\subset \{(Y_{1}, X_{2}): y_{1}\in \supp \psi, \ x_{2}\in \supp \chi, \ Y_{1}\sim X_{2}\}$, 
 
where the notation $Y\sim X$ is defined in (\ref{e8.6b}).
\item $\psi \rho\circ E \chi: \cD(M)\to \cD(\tC)$ extends continuously as $\psi \rho\circ E \chi: \cD'(M)\to \cD'(\tC)$.
\een
\end{proposition}
\proof
  It is well-known that:
\beq\label{e8.7}
\begin{array}{rl}
&\supp E\subset \{(x_{1}, x_{2}): \ x_{1}\in J(x_{2})\},\\[2mm]
&\WF(E)'=\{(X_{1}, X_{2})\in \cN\times \cN: \ X_{1}\sim X_{2}\}.
\end{array}
\eeq
On the other hand the distributional kernel  of $\rho$ equals 
\[
 \delta(r_{2})\otimes \delta(s_{1}, y_{1}, s_{2 },y_{2})\beta^{-1}(s_{1}, y_{1})\in \cD'(\tC\times M).
 \] It follows that: 
\beq\label{e8.8}
\begin{aligned}
\WF(\rho)'=\{ &(Y_{1}, X_{2}): \ r_{2}=0, \ (s_{1}, y_{1})= (s_{2}, y_{2}),  \\ &  (\sigma_{1}, \eta_{1})= (\sigma_{2},\eta_{2}), \  (\sigma_{2}, \eta_{2})\neq (0, 0)\}.
\end{aligned}
\eeq
  Since $E: \cD(M)\to \cE(M)$ we see that $\psi \rho\circ E \chi: \cD(M)\to \cD(\tC)$. Moreover there exists $\chi_{1}\in \coinf(M)$ such that  $\psi \rho\circ E \chi= \psi \rho\circ \chi_{1}E\chi$. 
  We have then
\[_{\tC}\WF(\rho)'=\WF(E)'_{M}=\emptyset,
\]
 and it follows from \cite[Chap. 8]{H1}  and (\ref{e8.7}), (\ref{e8.8}) that:
\[
\begin{aligned}
&\WF(\psi\rho\circ E \chi)'\subset \WF(\psi \rho)'\circ \WF(E \chi)'\\[2mm]
&\subset \{(Y_{1}, X_{2}): \exists\ \varrho_{1}\hbox{ s.t. }((0, s_{1}, y_{1}), (\varrho_{1}, \sigma_{1}, \eta_{1}))\sim X_{2}, \ x_{2}\in \supp \chi\}.
\end{aligned}
\]
Using  that $\supp \chi\cap C=\emptyset$ and Lemma \ref{8.2b} (1), this implies (1).
  Moreover (1) implies that 
 \beq\label{etiriti}
 \WF(\psi\rho\circ E \chi)'_{M}= \emptyset.
\eeq 
 Again by \cite{H1} this implies that $\psi \rho\circ E \chi= \cD(M)\to \cD(\tC)$ extends continuously as $\psi\rho\circ E \chi: \cD'(M)\to \cD'(\tC)$. \qed

\subsection{Hadamard condition on the cone}\label{sec8.2}
Recall from Subsect. \ref{butobound} that to a quasi-free state $\omega_{C}$ on $\CCR(\cH(\tC), \sigma_{C})$  we  can associate  a quasi-free state $\omega_{0}$ on 
$\CCR(\coinf(M_{0})/P\coinf(M_{0}), E_{0})$.
  In this subsection we give natural conditions on the covariances $\lambda^{\pm}$ of $\omega_{C}$ which ensure that the induced state $\omega_{0}$  satisfies the microlocal spectrum condition (\ref{musc}).

Recall that we denote by $Y= ((s,y), (\sigma, \eta))$ the points in $T^{*}\tC$. We also denote by $\Delta$ the diagonal in $T^{*}\tC\times T^{*}\tC$, {\Red and we will use the notation $_{\tC}\Gamma$, $\Gamma_{\tC}$ introduced  in \eqref{eq:newnotation}}.

\begin{theoreme}\label{th8.1}
Let $\lambda^{\pm}: \cH(\tC)\to \cH(\tC)$ and
\[
\Lambda^{\pm}\defeq  (\rho\circ E_{0})^{*}\circ \lambda^{\pm} \circ (\rho\circ E_{0}).
\]Then:
 \ben
 \item $\Lambda^{\pm}\in \cD'(M_{0}\times M_{0})$.
 
 \item If
 \[
\begin{array}{l}
i)\ \WF(\lambda^{\pm})'\cap \{(Y_{1}, Y_{2}): \pm\sigma_{1}<0\hbox{ or } \pm\sigma_{2}<0\}= \emptyset,\\[2mm]
ii)\  \WF(\lambda^{+}- \lambda^{-})'\cap \{(Y_{1}, Y_{2}): \sigma_{1}\hbox{  and }\sigma_{2}\neq 0\}\subset \Delta,
\end{array}
\]
then:
\[
iii)  \ \WF(\lambda^{\pm})'\cap \{(Y_{1}, Y_{2}): \pm\sigma_{1}>0\hbox{ and }\pm\sigma_{2}>0\}\subset \Delta.
\]
 
\item Assume moreover that $\lambda^{\pm}:\cH(\tC)\to \cH(\tC)$, $_{\tC}\!\WF(\lambda^{\pm})'= \WF(\lambda^\pm)'_{\tC}=\emptyset$. Then if  i) and iii)  in  (2) hold, 
 $\Lambda^{\pm}$ satisfy $(\mu\rm{sc})$.
\een
\end{theoreme}

\proof To prove (1) it suffices to check that $\rho\circ E_{0}: \cD(M_{0})\to \cH(\tC)$. If $\chi\in \coinf(M_{0})$, then  by Lemma \ref{6.-1} $\rho\circ E_{0}\chi=  \rho\circ\chi_{1} E \chi$ for some $ \chi_{1}\in \coinf(M)$.  Since  $ E: \cD(M)\to \cE(M)$ and $\rho: \cD(M)\to \cH(\tC)$ are continuous, this proves (1).

To  prove (2) we write:
\[
\begin{aligned}
&\WF(\lambda^\pm)'\cap \{\pm\sigma_1>0,\,\pm\sigma_2>0 \}  \\[2mm]
&\subset\big(\WF(\lambda^\mp)'\cap \{\pm\sigma_1>0,\,\pm\sigma_2>0 \}\big)\cup\big( \WF(\lambda^+-\lambda^-)'\cap  \{\pm\sigma_1>0,\,\pm\sigma_2>0 \}\big)\\[2mm]
&\subset\big(\WF(\lambda^\mp)'\cap \{\pm\sigma_1>0,\,\pm\sigma_2>0 \}\big)\cup\big( \WF(\lambda^+-\lambda^-)'\cap \{\sigma_1, \sigma_{2}\neq 0\}\big).
\end{aligned}
\]
The first set in the last line is empty by {\it i)}, and the second is  contained in $\Delta$ by {\it  ii)}.

To prove (3) we follow an argument due to Moretti \cite{Mo2}. We treat only the case of $\lambda^{+}$, the case of $\lambda^{-}$ being similar, and omit the $+$ superscript. Let $\chi_{i}\in \coinf(M_{0})$, $i=1,2$. By Prop. \ref{6.-00} there exists $\psi_{i}\in \coinf(C)$ (and hence $\psi_{i}\equiv 0$ near $p$) such that  any null geodesic starting from $\supp \chi_{i}$ intersects $C$ in $\{\psi_{i}=1\}$.
We have:
\[
\begin{aligned}
\chi_{1}\Lambda \chi_{2}=\,&  \chi_{1}(\rho\circ E)^{*}\psi_{1}\circ \lambda \circ \psi_{2}(\rho\circ E)\chi_{2}\\[2mm]
&+  \chi_{1}(\rho\circ E)^{*}\psi_{1}\circ \lambda \circ (1-\psi_{2})(\rho\circ E)\chi_{2}\\[2mm]
 &+ \chi_{1}(\rho\circ E)^{*}(1-\psi_{1})\circ \lambda \circ \psi_{2}(\rho\circ E)\chi_{2}\\[2mm]
   &+ \chi_{1}(\rho\circ E)^{*}(1-\psi_{1})\circ \lambda \circ(1- \psi_{2})(\rho\circ E)\chi_{2}\\[2mm]
   \eqdef\, & \Lambda_{1}+ \Lambda_{2}+ \Lambda_{3}+ \Lambda_{4}.
\end{aligned}
\]
By the properties of $\chi_{i}$, $\psi_{i}$, we can find $\tilde{\chi}_{i}\in \coinf(M)$ supported near $p$ such that: 
\[
\begin{array}{rl}
(a)&(1- \psi_{i}) (\rho\circ E) \chi_{i}= (1- \psi_{i}) \rho\circ \tilde{\chi}_{i} E \chi_{i},\\[2mm]
(b)&\hbox{ no null geodesic from supp}\chi_{i}\hbox{ intersects supp}\tilde{\chi}_{i}.
\end{array}
\]
It follows from ($b$)  and (\ref{e8.7}) that $\tilde{\chi}_{i}E \chi_{i}$ has a smooth compactly supported kernel, hence $\tilde{\chi}_{i}E \chi_{i}: \cD'(M)\to \cD(M)$.  Since $(1-\psi_{i})\rho:\cD(M)\to \cH(\tC)$ we see that
\begin{equation}
\label{e9.2}
 (1- \psi_{i}) \rho\circ E \chi_{i}: \cD'(M)\to \cH(\tC),
\end{equation}
hence
\begin{equation}
\label{e9.3}
\chi_{i}(\rho\circ E)^{*}(1- \psi_{i}): \cH'(\tC) \to \cD(M).
\end{equation}
It remains to examine the properties of $ \psi_{i} (\rho\circ E)\chi_{i}$.  By Prop. \ref{8.2} we know that $\psi_{i}(\rho\circ E)\chi_{i}: \cD'(M)\to \cE'(\tC)$. Since  $\cE'(\tC)\subset \cH'(\tC)$ continuously, we have
\beq\label{e9.4}
\psi_{i}(\rho\circ E)\chi_{i}: \cD'(M)\to \cH'(\tC),
\eeq
hence:
\begin{equation}
\label{e9.5}
\chi_{i}(\rho\circ E)^{*}\psi_{i}: \cH(\tC) \to \cD(M).
\end{equation}
From (\ref{e9.2}), \dots , (\ref{e9.5}) and the assumption that $\lambda: \cH(\tC)\to \cH(\tC)$ it follows that $ \Lambda_{i}: \cD'(M_{0})\to \cD(M_{0})$ hence has a smooth kernel for  $i=2,3,4$, and $\WF(\chi_{1}\Lambda \chi_{2})'= \WF(\Lambda_{1})'$.

To bound $\WF(\Lambda_{1})'$ we choose $\tilde{\psi}_{i}\in \coinf(\tC)$ such that $\tilde{\psi}_{i}\psi_{i}= \psi_{i}$ and write 
\[
\Lambda_{1}= \big(\chi_{1}(\rho\circ E)\psi_{1}\big)\circ \big(\tilde{\psi}_{1}\lambda \tilde{\psi}_{2}\big)\circ \big(\psi_{2}(\rho\circ E)\chi_{2})\big)\eqdef  K_{1}^{*}\circ d\circ K_{2},
\]
where $K_{i}= \psi_{i}(\rho\circ E)\chi_{i}\in \cE'(M\times \tC)$, $d= \tilde{\psi}_{1}c \tilde{\psi}_{2}\in \cE'(\tC\times \tC)$. The distributions $K_{1}$, $K_{2}$ and $d$ have compact support. Moreover we have
\[
\WF(d)'_{\tC}= _{\tC}\!\WF(d)'= \WF(K_{1})'_{M}= _{M}\!\WF(K_{2}^{*})'=\emptyset.
\]
In fact the first two equalities follow from the corresponding hypothesis on $\WF(c)'$, the last two  from (\ref{etiriti}). We can then apply  the results in \cite[Chap. 8]{H1} on the composition of kernels, and obtain that $K_{2}^{*}\circ d\circ K_{1}$ is well defined and
\[
\WF(K_{2}^{*}\circ d\circ K_{1})\subset \WF(K_{2}^{*})'\circ \WF(d)'\circ \WF(K_{1})'.
\]
Now we apply Prop. \ref{8.2} (1), the fact that $\WF(d)'\subset \WF(\lambda)'$ and Lemma \ref{8.2b} (1). We obtain that if $(X_{1}, X_{2})\in\WF(\Lambda)'$, necessarily $X_{1}, X_{2}\in \cN_{+}$ and  $X_{1}\sim X_{2}$, which is exactly  condition $(\mu\rm{sc})$. \qed

\section{Pseudodifferential calculus}\label{sec20}\init 
In this section we collect rather standard results on the pseudodifferential calculus on $\tC= \rr\times \bS$. We will however need to consider {\em bi-homogeneous} symbols on $\rr\times\bS$, i.e. symbols having different homogeneities in the covariables $\sigma$ and $\eta$, dual to $s$ and $\theta$. 

The reason for this is that the charge $q= -2 D_{s}$ is not an elliptic differential operator in the usual sense (considered on $\tilde C$), hence operators like $(q-z)^{-1}$ for $z\in \cc\backslash \rr$ are not in the usual pseudodifferential classes.

For $k, k'\in \rr$ we denote by $H^{k}(\rr)$,  $H^{k'}(\bS)$ the Sobolev space on $\rr$,  $\bS$ or order  $k$, $k'$, and by $\| \cdot\|_{k}$, $\| \cdot\|_{k'}$ their respective norms. Furthermore, we denote by $H^{k, k'}(\rr\times\bS)$  the Sobolev space on $\rr\times\bS$ of bi-order $(k, k')$, i.e. the completion of $C_0^\infty(\rr\times\bS)$ for the norm
\[
\| \psi\|_{k,k'}\defeq \| \langle D_s \rangle^k \langle D_\theta \rangle^{k'} \psi \|_2.
\]
We set also for $p\in\rr$:  
\[
B^{p}(\rr)= \bigcap_{k\in \rr}B(H^{k}(\rr), H^{k-p}(\rr)),
\]
equipped with its natural topology. 

\subsection{Pseudodifferential operators on $\rr\times \rr^{d-1}$}\label{sec20.1}
\begin{definition}
Let  $p_{1}, p_{2}\in\rr$. \ben
\item  we denote by $S^{p_{1}, p_{2}}(\rr\times \rr^{d-1})$  the space of  symbols $a\in \cinf(T^{*}\rr\times T^{*}\rr^{d-1})$ such that
 \[
|\p_{s}^{\alpha_{1}}\p_{\sigma}^{\beta_{1}}\p_{y}^{\alpha_{2}}\p_{\eta}^{\beta_{2}}a|\in O(\langle \sigma\rangle^{p_{1}- |\beta_{1}|}\langle \eta\rangle^{p_{2}- | \beta_{2}|}), \ \alpha_{1}, \beta_{1}\in \nn, \ \alpha_{2}, \beta_{2}\in \nn^{d-1}.
\]
\item we denote by $B^{p_{1}}S^{ p_{2}}(\rr\times\bS)$ the space of $a\in \cinf(T^{*}\rr^{d-1}, B^{p_{1}}(\rr))$ such that:
\[
\|\p_{y}^{\alpha_{2}}\p_{\eta}^{\beta_{2}}a\|_{p_{1}, k_{1}}\in O(\langle \eta\rangle^{p_{2}- |\beta_{2}|}), \ \alpha_{2}, \beta_{2}\in \nn^{d-1},
\]
where $\| \cdot \|_{p_{1}, k_{1}}$ is any seminorm of $a$ in $B^{p_{1}}(\rr)$.
\een
\end{definition}
Using the Weyl quantization on $\rr\times\rr^{d-1}$, we obtain a map
\[
S^{p_{1}, p_{2}}(\rr\times \rr^{d-1})\ni a \mapsto \Op(a)\in B(\coinf(\rr\times \rr^{d-1}), \cinf(\rr\times \rr^{d-1})),
\]
whose range, denoted by  $\Psi^{p_{1}, p_{2}}(\rr\times \rr^{d-1})$ is the space of  pseudodifferential operators on $\rr\times \rr^{d-1}$ of bi-order $(p_{1}, p_{2})$. Similarly using the Weyl quantization on $\rr^{d-1}$ we obtain a map 
\[
B^{p_{1}}S^{ p_{2}}(\rr\times \rr^{d-1})\ni a \mapsto \Op(a)\in B(\coinf(\rr\times \rr^{d-1}), \cinf(\rr\times \rr^{d-1})),
\]
whose range will be denoted by $B^{p_{1}}\Psi^{p_{2}}(\rr\times \rr^{d-1})$. 
\subsection{Pseudodifferential operators on $\tC$}\label{sec20.2}

Let   $A: \coinf(\tC)\to \cinf(\tC)$. If $\chi_{i}\in \cinf(\bS)$, $i=1,2$ are cutoff functions supported in chart open sets $\Omega_{i}\subset \bS$ and $\phi_{i}: \Omega_{i}\to \rr^{d-1}$ are coordinate charts, then $\phi_{1}^{*}\circ \chi_{1}A\chi_{2}\circ \phi_{2}^{-1*}: \coinf(\rr\times \rr^{d-1})\to \cinf(\rr\times \rr^{d-1})$.

\begin{definition}\label{defino}
\ben 
 \item We denote by $\Psi^{p_{1}, p_{2}}(\tC)$ the space of operators $A: \coinf(\tC)\to \cinf(\tC)$ such that for any $\chi_{i}$, $\phi_{i}$ as above $\phi_{1}^{*}\circ \chi_{1}A\chi_{2}\circ \phi_{2}^{-1*}\in \Psi^{p_{1}, p_{2}}(\rr\times \rr^{d-1})$.
 \item We denote by $B^{p_{1}}\Psi^{p_{2}}(\tC)$ the space of operators $A: \coinf(\tC)\to \cinf(\tC)$ such that for any $\chi_{i}$, $\phi_{i}$ as above $\phi_{1}^{*}\circ \chi_{1}A\chi_{2}\circ \phi_{2}^{-1*}\in B^{p_{1}}\Psi^{p_{2}}(\rr\times \rr^{d-1})$.
 \item We set 
 \[
\Psi^{-\infty, p_{2}}(\tC)= \bigcap_{p_{1}\in \rr} \Psi^{p_{1}, p_{2}}(\tC), \  B^{-\infty}\Psi^{p_{2}}(\tC)= \bigcap_{p_{1}\in \rr}B^{p_{1}}\Psi^{p_{2}}(\tC).
\]
\item We set
\[
\tilde\Psi^{p_{1}, p_{2}}(\tC)= \Psi^{p_{1}, p_{2}}(\tC)+ B^{-\infty}\Psi^{p_{2}}(\tC).
\]
  \een
\end{definition}
Note that if one defines analogously $\tilde{\Psi}^{-\infty, p_{2}}(\tC)\defeq \bigcap_{p_{1}\in \rr} \tilde{\Psi}^{p_{1}, p_{2}}(\tC)$, then actually $\tilde{\Psi}^{-\infty, p_{2}}(\tC)=B^{-\infty}\Psi^{p_{2}}(\tC)$. Moreover  it is easy to check that
\[
\tilde{\Psi}^{p_{1}, p_{2}}(\tC)\circ \tilde{\Psi}^{q_{1}, q_{2}}(\tC)\subset \tilde{\Psi}^{p_{1}+ p_{2}, q_{1}+ q_{2}}(\tC).
\]
We refer the reader to \cite{Rod, BS, RT} and references therein for more details on the pseudo-differential calculus on products of manifolds\footnote{Note however that the literature discusses mostly the case when both manifolds are compact.}.
\subsection{Beals criterion}\label{beals}
Let us denote by  $\Psi^{p}(\bS)$ the classes of standard pseudodifferential operators on $\bS$. It is well-known that $\Psi^{p}(\bS)$ can be characterized by the Beals criterion, namely an operator $A: \cinf(\bS)\to \cinf(\bS)$ belongs to $\Psi^{p}(\bS)$ iff 
\beq\label{multicam}
\ad_{f_{1}}\cdot\cdot\  \ad_{f_{n}}\ad_{X_{1}}\cdot\cdot\  \ad_{X_{m}}A: H^{k}(\bS)\to H^{k-p+n}(\bS), \ n, m\in \nn, \ k\in \zz,
\eeq
for any $f_{i}\in \cinf(\bS)$ and $X_{j}$ smooth vector fields on $\bS$ \cite{RT}. Moreover  one can find a finite set of such $f_{i}$ and $X_{j}$ such that the topology on $\Psi^{p}(\bS)$ given by the collection of the norms of the multi-commutators is equivalent to the standard topology on $\Psi^{p}(\bS)$, given by the 
symbol space topologies of the pullbacks $\phi_{i}^{*}\circ \chi_{i} A \chi_{j}\circ \phi_{j}$ in Def. \ref{defino}, for a fixed covering of $\bS$ by chart neighborhoods $U_{i}$.

These characterizations immediately carry over to the classes $B^{p_{1}}\Psi^{p_{2}}(\tC)$. In fact it is easy to see that $A\in B^{p_{1}}S^{p_{2}}(\tC)$ iff 
\beq\label{multicomm}
\ad_{f_{1}}\cdot\cdot\  \ad_{f_{n}}\ad_{X_{1}}\cdot\cdot\  \ad_{X_{m}}A: H^{k, k'}(\tC)\to H^{k-p_{1}, k'-p_{2}+n}(\bS), \ n, m\in \nn, k, k'\in \zz.
\eeq
This result can be deduced from the previous one by considering the operators 
\[
\left((u_{1}|\otimes\one_{\bS}\right)\circ A\circ \left(|u_{2})\otimes \one_{\bS}\right): \cinf(\bS)\to \cinf(\bS)
\]
for $u_{1}\in H^{-k+p_{1}}(\rr)$, $u_{2}\in H^{k}(\rr)$, which belongs to $\Psi^{p_{2}}(\bS)$ if (\ref{multicomm}) holds.   Applying the result recalled above about the equivalence of the standard topology and the topology given by the multicommutator norms, one obtains that $A\in B^{p_{1}}\Psi^{p_{2}}(\tC) $ if (\ref{multicomm}) holds.

In the usual case one can deduce from the Beals criterion standard results on the functional calculus for pseudo-differential operators, for example on complex powers of elliptic $\Psi$DOs \cite{Bo}. These results are easy to extend to the classes $B^{p_{1}}\Psi^{p_{2}}(\tC)$. We will need only a very simple one, which we now state. Recall that $\tilde\Psi^{-\infty, 0}(\tC)= B^{-\infty}\Psi^{0}(\tC)\subset B(L^{2}(\tC))$. The spectrum of $b\in B(L^{2}(\tC))$ is denoted ${\rm spec}(b)$.

\begin{proposition}\label{beals-prop}
 Let $b\in \tilde\Psi^{-\infty, 0}(\tC)$ and  $F$ holomorphic near ${\rm spec}(b)$ with  $F(0)=0$. Then $F(b)\in \tilde\Psi^{-\infty, 0}(\tC)$.
\end{proposition}
\proof The proof consists of expressing $F(b)$ as a contour integral and applying  the Beals criterion  to the resolvent $(b-z)^{-1}$. \qed
\subsection{Essential support}
We denote by $\Psi^{p}_{\rm ph}(\rr)$, $p\in \rr$ the class of global  pseudo-differential operators on $\rr$ with poly-homogeneous symbols.
\begin{definition}\label{normal}
 The {\em essential support} of $a\in \Psi^{p_{1}, p_{2}}(\tC)$, denoted by $
 \esssupp(a)\subset T^{*}\rr\backslash Z$
 is defined by:
 
   $(s_{0}, \sigma_{0})\not\in {\esssupp}(a)$ if there exists $b\in \Psi^{0}_{\rm ph}(\rr)$, elliptic at $(s_{0}, \sigma_{0})$ such that 
$b\circ a\in \Psi^{-\infty, p_{2}}(\tC)$. 
\end{definition}
Clearly ${\esssupp}(a)$ is a closed conic subset of $T^{*}\rr\backslash Z$. Moreover one can equivalently require that $a\circ b \in \Psi^{-\infty, p_{2}}(\tC)$ for some $b\in \Psi^{0}_{\rm ph}(\rr)$, elliptic at $(s_{0}, \sigma_{0})$. 

\subsection{Wavefront set of kernels}
 For $N= \rr, \bS, \rr\times\bS$, we denote by $\Delta_{N}$ the diagonal in $T^{*}N\times T^{*}N$,  and by $Z_{N}$ the zero section in $T^{*}N$.
 
For an operator $a\in \Psi^{p_{1}, p_{2}}(\rr\times \bS)$ it is in general not true that $\WF(a)'$ is contained in the full diagonal $\Delta_{\rr\times\bS}$ (as would be the case for an operator in $\Psi^p(\rr\times \bS)$). Instead one has the following estimate, which can be thought as a natural generalization of the usual estimate for the wave front set of tensor products of distributions (in this case Schwartz kernels) \cite{BS}. 
 
\begin{lemma}\label{schulz}
Let $a\in  \Psi^{p_{1}, p_{2}}(\rr\times \bS)$. Then: 
\[
\WF(a)'\subset \Delta_{\rr}\times \Delta_{\bS}\cup \Delta_{\rr}\times (Z_{\bS}\times Z_{\bS})\cup (Z_{\rr}\times Z_{\rr})\times  \Delta_{\bS}.
\]
\end{lemma}
\def\tPsi{\tilde{\Psi}}
\def\nlam{\langle \lambda\rangle}

Less precise estimates are valid for the $\tilde \Psi^{p_{1}, p_{2}}(\rr\times \bS)$ classes:

\begin{lemma}\label{schulz.bis}
\ben
\item Let $a\in B^{-\infty}\Psi^{p_{2}}(\tC)$. Then
\[
 \WF(a)'\cap \{(Y_{1}, Y_{2}): \sigma_{1}\neq 0\hbox{ or }\sigma_{2}\neq 0\}= \emptyset.
\]
 \item Let $a\in \tPsi^{p_{1}, p_{2}}(\tC)$. Then
 \[
 _{\tC}\WF(a)'= \WF(a)'_{\tC}= \emptyset.
\]
\een
\end{lemma}
The proof is given in Subsect. \ref{A2}.
\subsection{Toeplitz pseudo-differential operators on $\tC$}\label{top}
We recall that $\cH(\tC)= \bigcap_{m\in \rr}H^{m}(\tC)= \bigcap_{k\in \rr}H^{k, k}(\tC).$
Let us set
\[
L^{2}_{\pm}(\tC)\defeq \one_{\rr^{\pm}}(D_{s})L^{2}(\tC)
\]
 and denote by $i_{\pm}: L^{2}_{\pm}(\tC)\to L^{2}(\tC)$ the corresponding isometric injection, so that $\pi_{\pm}\defeq  i_{\pm}i_{\pm}^{*}= \one_{\rr^{\pm}}(D_{s})$ is the orthogonal projection on $L^{2}_{\pm}(\tC)$ in $L^{2}(\tC)$. We set also
\beq\label{top.1}
\cH_{\pm}(\tC)\defeq  i^{*}_{\pm}\cH(\tC)\subset \cH(\tC).
\eeq
We will see in Sect. \ref{sec21} that this provides a useful setup for the discussion of the positivity condition $\lambda^\pm\geq 0$ for the two-point functions of a Hadamard state.

Writing $\one_{\rr^{\pm}}= \chi\one_{\rr^{\pm}}+ (1- \chi)\one_{\rr^{\pm}}$ for  a cutoff function $\chi\in \coinf(\rr)$ equal to $1$ near $0$, we see that 
\beq\label{top.0}
\pi_{\pm}\in \tPsi^{0, 0}(\tC).
\eeq
For $\alpha, \beta\in \{+, -\}$ and $p_{1}, p_{2}\in \rr$ we set:
\[
\tPsi^{p_{1}, p_{2}}_{\alpha\beta}(\tC)\defeq  i_{\alpha}\circ \tPsi^{p_{1}, p_{2}}(\tC)\circ i_{\beta}^{*}.
\]
By (\ref{top.1}) we see that $\tPsi^{p_{1}, p_{2}}_{\alpha\beta}(\tC): \cH_{\beta}(\tC)\to \cH_{\alpha}(\tC)$. Moreover  if we set:
\[
R_{\alpha\beta}: \  \tPsi^{p_{1}, p_{2}}(\tC)\ni a\mapsto i_{\alpha}^{*}\circ a\circ i_{\beta}\in \tPsi^{p_{1}, p_{2}}_{\alpha\beta}(\tC),
\]
then  using (\ref{top.0}), we see that $R_{\alpha\beta}$ has right inverse
\[
T_{\alpha\beta}:\tPsi^{p_{1}, p_{2}}_{\alpha\beta}(\tC)\ni a\mapsto i_{\alpha}\circ a\circ i_{\beta}^{*}\in 
\tPsi^{p_{1}, p_{2}}(\tC),
\]
which allows to identify $\tPsi^{p_{1}, p_{2}}_{\alpha\beta}(\tC)$ with $\Ran T_{\alpha\beta}\subset \tPsi^{p_{1}, p_{2}}(\tC)$.
From (\ref{top.0}) we also have:
\begin{equation}
\label{top.2}
\tPsi^{p_{1}, p_{2}}_{\alpha\beta}(\tC)\circ \tPsi^{q_{1}, q_{2}}_{\beta\gamma}(\tC)\subset \tPsi^{p_{1}+q_{1}, p_{2}+q_{2}}_{\alpha\gamma}(\tC).
\end{equation}

\def\tut{(2|D_{s}|)^{\12}}

\section{Construction of Hadamard states on the cone}\label{sec21}
 From the discussion in Subsect. \ref{sec8.2}, in particular Thm. \ref{th8.1}, we are led to the following definition.
 \begin{definition}\label{def21.1}
 A pair of maps $\lambda^{\pm}: \cH(\tC)\to \cH(\tC)$ is called a pair of {\em  Hadamard two-point functions} on the cone $C$ if:
 \beq\label{hadada}\tag{Had}
\begin{array}{rl}
i)& _{\tC}\WF(\lambda^{\pm})'= \WF(\lambda^{\pm})'_{\tC}= \emptyset,\\[2mm]
ii)&\WF(\lambda^{\pm})'\cap \{(Y_{1}, Y_{2}): \pm \sigma_{1}<0\hbox{ or }\pm \sigma_{2}<0\}= \emptyset,\\[2mm]
iii)& \lambda^{+}- \lambda^{-}= 2 D_{s},\\[2mm]
iv)& \lambda^{\pm}\geq 0 \hbox{ on }\cH(\tC). 
\end{array}
\eeq
\end{definition}

As the name suggests, if $\lambda^{\pm}$ are Hadamard two-point functions on $C$ in the sense of the above definition, then $\Lambda^{\pm}$ defined in (\ref{defdeLam}) are Hadamard two-point functions on $M_{0}$ (as follows from Thm. \ref{th8.1}).

We now discuss in more detail the various conditions in (\ref{hadada}). It is natural to consider  pseudodifferential  two-point functions, i.e. to assume that  $\lambda^{\pm}\in \tPsi^{p_{1}, p_{2}}(\tC)$.
Moreover to analyze conditions (\ref{hadada}) {\it iii)}, {\it iv)} it is convenient to reduce oneself to $\lambda^{\pm}$ of the form:
\begin{equation}
\label{form}
\lambda^{\pm}= \tut c^{\pm}\tut, \hbox{ where }c^{\pm}\in \tPsi^{p_{1}, p_{2}}(\tC),
\end{equation}
for $p_{1}, p_{2}\in \rr$.   Note  that writing 
$\tut$ as $\chi(D_{s})\tut+ (1- \chi(D_{s}))\tut$ for $\chi\in \coinf(\rr)$ equal to $1$ near $0$,  we see that (\ref{form}) implies that $\lambda^{\pm}\in \tPsi^{p_{1}+1, p_{2}}(\tC)$.

\subsection{Wavefront set}
We first analyze conditions  (\ref{hadada}) {\it i)},  {\it ii)}.
\begin{proposition}\label{21.2}
Assume that \beq
\label{tour}
\begin{array}{rl}
&\lambda^{\pm}= a^{\pm}+ r^{\pm}, \ a^{\pm}\in \Psi^{p_{1}, p_{2}}(\tC),\ r^{\pm}\in \tilde{\Psi}^{-\infty, p_{2}}(\tC),\\[2mm]
& (\rr\times \rr^{\mp})\cap {\esssupp}(a^{\pm})= \emptyset.
\end{array}
\eeq
 Then $\lambda^{\pm}$ satisfies conditions   (\ref{hadada}) {\it i)}, {\it ii)}.
 \end{proposition}
\proof
The fact that $\lambda^{\pm}$ satisfy {\it i)} follows from Lemma \ref{schulz.bis} (2). Also since by Lemma \ref{schulz.bis} (1) $r^{\pm}$ satisfy {\it ii)} we can assume that $\lambda^{\pm}= a^{\pm}$. We treat only the case of $\lambda^{+}$ and use  the notation in the proof of Lemma \ref{schulz.bis}. Let $\tilde{Y}_{1}, \tilde{Y}_{2}\in T^{*}\tC\backslash Z$ with $\tilde{\sigma}_{1}\neq 0$ or $\tilde{\sigma}_{2}\neq 0$.  Let us assume that $\tilde{\sigma}_{1}\neq 0$, the case $\tilde{\sigma}_{2}\neq 0$ being similar, using the remark after Def. \ref{normal}.

Since $(\rr\times \rr^{+})\cap \,{\esssupp}(a^{+})= \emptyset$, we can find  a cutoff function $\chi_{1}$ with $\chi_{1}(\tilde{s}_{1})\neq 0$, a neighborhood $V_{1}$ of $\tilde{\sigma}_{1}$ and some $m_{1}\in \Psi^{0}_{\rm ph}(\rr)$ elliptic at $(\tilde{s}_{1}, \tilde{\sigma}_{1})$ such that $(1-m_{1})(s, D_{s})v_{\sigma, \lambda}\in O(\nlam^{-\infty})$ in all $H^{k}(\rr)$ and $m_{1}(s, D_{s})\circ a\in \tPsi^{-\infty, p_{2}}(\tC)$. The fact that $(\tilde{Y}_{1}, \tilde{Y}_{2})\not\in \WF(a)'$ follows then from the same arguments as in the proof of Lemma \ref{schulz.bis}. \qed

\medskip

In terms of $c^{\pm}$ appearing in (\ref{form}), a  natural condition implying (\ref{tour}) is
\begin{equation}
\label{muscc}\tag{$\mu{\rm sc}_{C}$}
 \one_{\rr^{\mp}}(D_{s})c^{\pm}\in \tPsi^{-\infty, p_{2}}(\tC),
\end{equation}
which clearly implies  that $\lambda^{\pm}$ satisfy (\ref{tour}).
\begin{lemma}\label{21.3}
 Let $\lambda^{\pm}$ be given by (\ref{form}) such that  (\ref{muscc}) holds.  Then
 \[
c^{\pm}= \one_{\rr^{\pm}}(D_{s})+ \tPsi^{-\infty, p_{2}}(\tC).
\]
\end{lemma}
\proof In terms of $c^{\pm}$ (\ref{hadada}) {\it iii)} becomes $c^{+}- c^{-}=  \sgn(D_{s})$.  Let $\chi^{\pm}\in \cinf(\rr)$ be cutoff functions equal to $1$ near $\pm\infty$ and to $0$ near $\mp \infty$. From condition $(\mu{\rm sc})_{C}$ and pseudodifferential calculus we obtain that
\beq\label{21.e2}
 c^{\pm}= \chi^{\pm}(D_{s})c^{\pm}\chi^{\mp}(D_{s}) + \tPsi^{-\infty, p_{2}}(\tC).
\eeq
Using successively (\ref{21.e2}) and $c^{+}- c^{-}=  \sgn(D_{s})$ we obtain 
\[
\begin{aligned}
c^{\pm}&= \chi^{\pm}c^{\pm}\chi^{\pm}+ \tPsi^{-\infty, p_{2}}(\tC)\\[2mm]
&=\chi^{\pm}( c^{\mp}\pm \sgn(D_{s}))\chi^{\pm}+ \tPsi^{-\infty, p_{2}}(\tC)\\[2mm]
&=\chi^{\pm}c^{\mp}\chi^{\pm} + \chi^{\pm} \chi^{\pm} + \tPsi^{-\infty, p_{2}}(\tC)\\[2mm]
&=\chi^{\pm}\chi^{\mp}c^{\mp}\chi^{\mp}\chi^{\pm} +  \chi^{\pm} + \tPsi^{-\infty, p_{2}}(\tC)\\[2mm]
&=  \chi^{\pm}+  \tPsi^{-\infty, p_{2}}(\tC)\\[2mm]
&=\one_{\rr^{\pm}}(D_{s})+  \tPsi^{-\infty, p_{2}}(\tC). \ \ \Box
\end{aligned}
\]

\subsection{Positivity}
We now discuss conditions (\ref{hadada}) {\it iii)}, {\it iv)}.
In terms of $c^{\pm}$ they  become:
\begin{equation}
\label{21.e4}
\begin{array}{rl}
iii)& c^{+} -c^{-}= \sgn(D_{s}),\\[2mm]
iv)& c^{\pm}\geq 0\hbox{ on }\cH(\tC).
\end{array}
\end{equation}
To analyze (\ref{21.e4}) we use the framework of Subsect. \ref{top}. We denote $c^{+}$ simply by $c$ and set 
\[
c_{\alpha\beta}= i_{\alpha}^{*}\circ c\circ i_{\beta}, \ \alpha, \beta\in \{+,-\},
\]
so that:
\beq\label{formc}
c= \sum_{\alpha, \beta\in \{+,-\}}i_{\alpha}c_{\alpha\beta}i_{\beta}^{*}.
\eeq
Then (\ref{21.e4}) is equivalent to:
\begin{equation}
\label{21.e5}
\mat{c_{++}}{c_{+-}}{c_{-+}}{c_{--}}\geq 0, \ \mat{c_{++}- \one}{c_{+-}}{c_{-+}}{c_{--}+\one}\geq 0 \hbox{ on }\cH_{+}(\tC)\oplus \cH_{-}(\tC),
\end{equation}
which is equivalent to: 
\[
\begin{array}{rl}
i)& c_{++}\geq 0, \ c_{--}\geq \one, \ c_{-+}= c_{+-}^{*},\\[2mm]
ii)&\begin{array}{l}
|(u_{+}| c_{+-}u_{-})|\leq (u_{+}|c_{++}u_{+})^{\12}(u_{-}| c_{--}u_{-})^{\12},
\\
|(u_{+}| c_{+-}u_{-})|\leq (u_{+}|(c_{++}-\one)u_{+})^{\12}(u_{-}|( c_{--}+\one)u_{-})^{\12}, \ \ u_{\pm}\in \cH_{\pm}(\tC). \end{array}
\end{array}
\]
Condition {\it ii)} above is implied by
\[
|(u_{+}| c_{+-}u_{-})|\leq (u_{+}|(c_{++}-\one)u_{+})^{\12}(u_{-}|c_{--}u_{-})^{\12}, \ \ u_{\pm}\in \cH_{\pm}(\tC).
\]
We are now in position to prove the following theorem, which is the analog of \cite[Thm. 7.5]{GW} in the present situation. It provides a rather large class of Hadamard two-point functions on $C$, hence by Thm. \ref{th8.1}, of Hadamard states on $M_{0}$.
\begin{theoreme}\label{thm:constr}
 Assume that 
\[
\begin{array}{rl}
&c_{++}= \one + a_{+}^{*}a_{+}, \ c_{--}=a_{-}^{*}a_{-},\\[2mm]
&c_{+-}= c_{-+}^{*}= a_{+}^{*}d a_{-},
\end{array}
\]
for $a_{+}\in \tPsi_{++}^{-\infty, 0}(\tC)$, $a_{-}\in \tPsi_{--}^{-\infty, 0}(\tC)$, $d\in \tPsi^{0, 0}_{+-}(\tC)$ with $\|d\|_{B(L^{2}_{-}(\tC), L^{2}_{+}(\tC))}\leq \one$.

Let $c$ be given by (\ref{formc}) and $\lambda^{+}=\tut c\tut$, $\lambda^{-}= \lambda^{+}- 2 D_{s}$. Then $\lambda^{\pm}$ is  a pair of Hadamard two-point functions on the cone.
\end{theoreme}
\proof
We set as before $\lambda^{\pm}= \tut c^{\pm}\tut\in \tPsi^{1, 0}(\tC)$, so that $c^{+}=c$, $c^{-}= c- \sgn(D_{s})$. Condition (\ref{21.e4}) follows from the above discussion.  It remains to check condition (\ref{muscc}).  We embed   the spaces $\tPsi_{\alpha\beta}^{p_{1}, p_{2}}(\tC)$ into $\tPsi^{p_{1}, p_{2}}(\tC)$ as explained at the end of Subsect. \ref{top}, and we have:
\[
\begin{aligned}
c^{+}&=a_{+}^{*}a_{+}+ a_{+}^{*}da_{-} + a_{-}^{*}d^{*}a_{+}+ a_{-}^{*}a_{-}+ 
\one_{\rr^{+}}(D_{s}),\\[2mm]
c^{-}&= a_{+}^{*}a_{+}+ a_{+}^{*}da_{-} + a_{-}^{*}d^{*}a_{+}+ a_{-}^{*}a_{-}+ 
\one_{\rr^{-}}(D_{s}),
\end{aligned}
\]
hence
\[
\begin{aligned}
\one_{\rr^{-}}(D_{s})c^{+}&= a_{+}^{*}a_{+}+ a_{+}^{*}da_{-}\in \tPsi^{-\infty,0}(\tC), \\[2mm]
\one_{\rr^{+}}(D_{s})c^{-}&=a_{-}^{*}d^{*}a_{+}+ a_{-}^{*}a_{-}\in \tPsi^{-\infty, 0}(\tC),
\end{aligned}
\]
and condition (\ref{muscc}) is satisfied. \qed

\begin{remark}The special choice of vanishing $a_+$, $a_-$ and $d$ in Thm. \ref{thm:constr} gives two-point functions 
\[
\lambda^\pm = \pm 2 \one_{\rr^{\pm}}(D_{s}) D_{s}.
\]
In the setting of asymptotically flat spacetimes with past time  infinity $i^-$ these correspond to the Hadamard state found and further studied in \cite{Mo1,Mo2}.\end{remark}

\section{Pure Hadamard states}\label{sec22}\init
In this section we first characterize {\em pure} Hadamard states on the cone $C$. We then prove that any pure Hadamard state $\omega_{C}$ on $C$ induces a pure Hadamard state $\omega_{0}$ in $M_{0}$.

\subsection{An abstract criterion for purity}
Let $(\cY, \sigma)$ a complex symplectic space and $\omega$ a gauge invariant quasi-free state on $\CCR(\cY, \sigma)$, with complex covariances $\lambda^{\pm}$.  

Let $\cY^{\cpl}$ the completion of $\cY$ for the norm
\beq\label{def-de-norm}
\|y\|_{\omega}\defeq   (\overline{y}\cdot \lambda^{+}y+ \overline{y}\cdot\lambda^{-}y)^{\12}.
\eeq

Let us introduce the hermitian form $q= \i \sigma\in L_{\rm h}(\cY, \cY^{*})$. Clearly $q$, $\lambda^{\pm}$ extend uniquely to $\cY^{\cpl}$.  Then $\omega$ is pure  iff \cite{araki}:
\ben
\item $q$ is non-degenerate on $\cY^{\cpl}$, 
\item there exists an involution $\kappa:\cY^{\cpl}\to\cY^{\cpl}$ such that $\kappa^*q\kappa=q$, $q\kappa\geq 0$ and $\lambda^{\pm}= \12 q(\kappa\pm \one)$.
\een

From this discussion we obtain immediately the following lemma.
\begin{lemma}\label{lemlem}
Let $(\cY_{i}, \sigma_{i})$, $i=1, 2$ be two complex symplectic spaces and $\rho: \cY_{1}\to \cY_{2}$ an injective map such that $\rho^{*}\sigma_{2}\rho= \sigma_{1}$.  Let $\omega_{2}$ be a pure, gauge-invariant quasi-free state on $\CCR(\cY_{2}, \sigma_{2})$. Let $\omega_{1}$ the gauge invariant quasi-free  state on $\CCR(\cY_{1}, \sigma_{1})$ defined by the complex covariances 
\[
 \lambda_{1}^{\pm}= \rho^{*} \lambda_{2}^{\pm }\rho.
\]
 
Then if  $\rho\cY_{1}$ is dense in $\cY_{2}$ for the norm $\| \cdot\|_{\omega_{2}}$  defined in (\ref{def-de-norm}), 
  the state $\omega_{1}$ is pure on $\CCR(\cY_{1}, \sigma_{1})$.
\end{lemma}
\subsection{Pure Hadamard states on the cone} 
The following theorem is the exact analog of \cite[Thm. 7.10]{GW}. In what follows we will use the notations introduced in Subsect. \ref{top}.
\begin{theoreme}\label{thm6}
Let $\lambda^{\pm}$ be the two-point functions of a  state $\omega_{C}$ on $(\cH(\tC), \sigma_{C})$ of the form (\ref{form}) and satisfying (\ref{muscc}). Then $\omega_{C}$ is pure iff there exists $a\in \tPsi^{-\infty, 0}_{-+}(\tC)$ such that
\[
 c^{+}=  \mat{\one +a^{*}a}{a^{*}(\one+ aa^{*})^{\12}}{ (\one + aa^{*})^{\12}a}{aa^{*}}.
\]
\end{theoreme}
\proof 
We consider the pair $c^{\pm}$ obtained from $\lambda^{\pm}$, denote as before $c^{+}$ by $c$ and identify $c$ with the matrix $\mat{c_{++}}{c_{+-}}{c_{-+}}{c_{--}}$. Arguing as in the proof of \cite[Thm. 7.10]{GW}, we obtain that  the state $\omega_{C}$ on $(\cH(\tC), \sigma_{C})$ with covariances $\lambda^{\pm}$ is pure iff
\begin{equation}
\label{21.e10}
c=\mat{\one +a^{*}a}{a^{*}(\one+ aa^{*})^{\12}}{ (\one + aa^{*})^{\12}a}{aa^{*}},
\end{equation}
for some $a: L^{2}_{+}(\tC)\to L^{2}_{-}(\tC)$.   This proves $\Leftarrow$.

Let us now prove $\Rightarrow$.
Since we assumed that $c^{\pm}\in \tPsi^{0, 0}(\tC)$ satisfy (\ref{muscc}), we obtain that 
\begin{equation}
\label{21.e11}
a^{*}a\in \tPsi^{-\infty, 0}_{++}(\tC),  \ (\one + aa^{*})^{\12}a\in \tPsi^{-\infty, 0}_{-+}(\tC).
\end{equation}
We claim that 
\beq\label{claim}
(\one+ aa^{*})^{-\12}\in \one + \tPsi_{--}^{-\infty, 0}(\tC).
\eeq 
Let us prove (\ref{claim}). We use the operators $R_{\alpha\beta}$, $T_{\alpha\beta}$ defined  at the end of Subsect. \ref{top}.  We first embed $aa^{*}$ into $\tPsi^{-\infty, 0}(\tC)$, i.e. consider 
 $b= T_{--} (a^{*}a)$.  Then $b\geq 0$ on $L^{2}(\tC)$ and applying Prop. \ref{beals-prop} to $F(z)= (1+ z)^{\12}- 1$ we obtain that $(\one + b)^{-\12}-\one\in \tPsi^{-\infty, 0}(\tC)$. Writing $b$ as a $2\times 2$ matrix acting on $L^{2}_{+}(\tC)\oplus L^{2}_{-}(\tC)$ we see that $R_{++}\left((\one + b)^{\12}\right)= (\one + aa^{*})^{\12}$, which proves (\ref{claim}).  From (\ref{claim}) and (\ref{21.e11}) we obtain that $a\in \tPsi^{-\infty, 0}_{-+}(\tC)$. \qed

\medskip

In the next lemma we identify the completion of $\cH(\tC)$ for the norm (\ref{def-de-norm}) associated to any  Hadamard state considered in Thm. \ref{thm6}.

Let us first fix some  notation. For $a: L^{2}_{+}(\tC)\to L^{2}_{-}(\tC)$ we denote by $c^{+}(a)$ the operator defined in (\ref{21.e10}) set  $c^{-}(a)= c^{+}(a)- \sgn(D_{s})$, and \beq
\label{21.e12}\lambda^{\pm}(a)= \tut c^{\pm}(a)\tut.
\eeq 

If $\cH$ is a Hilbert space and $h\geq 0$ is a selfadjoint operator on $\cH$ with $\Ker h=\{0\}$, we denote by $h\cH$ the completion of $\Dom h^{-1}$ (i.e. the range of $h$) for the norm $\| h^{-1}u \|_{\cH}$.

\begin{lemma}\label{birmo}
 Let $a:  L^{2}_{+}(\tC)\to L^{2}_{-}(\tC)$. Then the completion of $\cH(\tC)$ for the norm $(\cdot| \left(\lambda^{+}(a)+ \lambda^{-}(a)\right)\cdot)^{\12}$ equals $|D_{s}|^{-\12}L^{2}(\tC)$.
\end{lemma}
\proof By (\ref{21.e12}) and the definition of $|D_{s}|^{-\12}L^{2}(\tC)$ it suffices to prove that the completion of $\cH(\tC)$ for the norm $(u| \left(c^{+}(a)+ c^{-}(a)\right)u)^{\12}$ equals $L^{2}(\tC)$.  Let 
\[
u(a)= \mat{(\one + aa^{*})^{\12}}{a}{a^{*}}{(\one + a^{*}a)^{\12}}, 
\]
and note that
\begin{equation}
\label{21.e13}
u(a)^{*}c^{\pm}(0)u(a)= c^{\pm}(a).
\end{equation}
Moreover using the  identity $af(a^{*}a)= f(aa^{*})a$, valid  for any Borel function $f$, we obtain that 
$u(a)^{-1}= u(-a)$, hence $u(a): L^{2}(\tC)\to L^{2}(\tC)$ is boundedly invertible.  By (\ref{21.e13}) it suffices to treat the case $a=0$ which is obvious since $c^{+}(0)+ c^{-}(0)= \one$. \qed

\subsection{Pure Hadamard states in $M_{0}$}\label{ss:pureint}

Our main result concerns the purity of the states induced in the bulk. We postpone the introduction of the key technical ingredients of the proof to Subsect. \ref{sec22.2} for the sake of self-consistency of our results on the characteristic Cauchy problem.

\begin{theoreme}\label{thm66}
Assume that $\dim M \geq 4$.  Let $\omega_{C}$ be a pure Hadamard state on $\CCR(\cH(\tC), \sigma_{C})$ as in Thm. \ref{thm6}. Then the state $\omega$ induced by $\omega_{C}$ on $\CCR(\coinf(M_{0})$ $/P\coinf(M_{0}), E_{0})$ is a pure state. \end{theoreme}
\proof The proof relies on Lemma \ref{lemlem} and on some results on the characteristic Cauchy problem in $M_{0}$, proved below in Subsect. \ref{sec22.2}. Recall that the map $\rho:\Sol(P_{0})\to \cH(\tC)$ was introduced in Def. \ref{10.1}.
By Lemmas \ref{lemlem} and \ref{birmo} it suffices to check that $\rho(\Sol(P_{0}))$ is dense in $| D_{s}|^{-\12}L^{2}(\tC)$. Since $\coinf(\rr\times \bS)$ is dense in $| D_{s}|^{-\12}L^{2}(\tC)$ it suffices for $w\in \coinf(\rr\times \bS)$ to find a sequence $\phi_{n}\in \Sol(P_{0})$ such that $\rho \phi_{n}\to w$ in $| D_{s}|^{-\12}L^{2}(\tC)$.

We will use freely the notation introduced below in Subsect. \ref{sec22.2}.
We  first fix a Cauchy surface $\Sigma$ in $(M, g)$ as in \ref{embedd} to the future of $\supp w$. Note that since $w$ vanishes near $s= -\infty$, we know that $w$ belongs to the space  $\tilde{H}_{0}^{1}(\tilde{C}_{0})$ introduced in Prop. \ref{propiprop}.  
By Thm. \ref{theothm} and Prop. \ref{propiprop}, there exists  $f$ in the energy space $\cE_{0}(\Sigma_{0})$ such that $w= R\circ T f$. Since $\coinf(\Sigma_{0})\oplus \coinf(\Sigma_{0})$ is dense in $\cE_{0}(\Sigma_{0})$, there exists a sequence $f_{n}\in \coinf(\Sigma_{0})\oplus \coinf(\Sigma_{0})$ such that $f_{n}\to f$ in $\cE_{0}(\Sigma_{0})$. By Thm. \ref{theothm} and Prop. \ref{propiprop}  we have $R\circ T f_{n}\to w$ in $\tilde{H}_{0}^{1}(\tilde{C}_{0})$, hence also $R\circ T f_{n}\to w$ in $|D_{s}|^{-\12}L^{2}(\tC)$, by Remark \ref{remrem}.

Let $\phi_{n}\in \Sol(P_{0})$ the solution with Cauchy data $f_{n}$ on $\Sigma_{0}$. Then  
$\rho \phi_{n}= R\circ T f_{n}\to w$  in $|D_{s}|^{-\12}L^{2}(\tC)$, which completes the proof of the theorem. \qed

\subsection{A characteristic Cauchy problem in $M_{0}$}\label{sec22.2}
\def\tSig{\tilde{\Sigma}}
From Lemma \ref{lemlem}, we see that to deduce purity of the bulk state from the purity of the boundary state,
the range of $\rho$ in $\cH(\tC)$ should be sufficiently large. One way to ensure this is to solve  a {\em characteristic Cauchy problem} in $M_{0}$, i.e. to construct an inverse for $\rho$.  If $M$ has a {\em compact} Cauchy surface, the characteristic problem was shown to be well posed in {\em energy spaces} by H\"{o}rmander \cite{Ho}. With some care the results of \cite{Ho} can be used in our situation.

\subsubsection{Characteristic Cauchy problem for compact Cauchy surfaces}\label{sec22.2.1}
We  recall an  important result of H\"{o}rmander \cite{Ho} on  the characteristic Cauchy problem in energy spaces.  The framework of \cite{Ho} is as follows: 

One considers a spacetime $(\tilde{M}, \tilde{g})$ for  $\tilde{M}= \rr\times \tilde{\Sigma}$,  $\tilde\Sigma$ a smooth {\em compact} manifold and $\tilde{g}= -\tilde{\beta}(t, \rx) dt^{2}+ \tilde{h}_{ij}(t, \rx)d\rx^{i}d\rx^{j}$.  One also fixes a real function $\tilde{r}\in \cinf(\tilde{M})$.
  
 If $\tSig_{1}$ is a Cauchy hypersurface in $(\tilde{M}, \tilde{g})$, we will denote by 
 \[
 \tilde{U}_{\tSig_{1}}: \cinf(\tilde{\Sigma}_{1})\oplus \cinf(\tilde{\Sigma}_{1})\to \cinf(\tilde{M})
 \]
  the Cauchy evolution operator for $-\Box_{\tilde{g}}+ \tilde{r}$, so that $\phi= \tilde{U}_{\tilde{\Sigma}_{1}}f$ solves
 \[
\begin{cases}
-\Box_{\tilde{g}}\phi+ \tilde{r}\phi=0, \\
\phi\tra{\tilde{\Sigma}_{1}}=f^{0}, \ n^{\mu} \nabla_{\mu}\phi\tra{\tilde{\Sigma}_{1}}= f^{1}.
\end{cases}
\]
 
 A hypersurface  $\tilde{C}$ of the form
 \beq\label{formdeC}
 \tilde{C}=\{(F(\rx), \rx) :\ \rx\in \tilde{\Sigma}\},   \ F\hbox{ Lipschitz},
 \eeq
  is called {\em space-like}  (resp. {\em weakly space-like}) if  
  \[
  \sup_{\rx\in \tilde{\Sigma}}\left(-\beta^{-1}(F(\rx), \rx)+\p_{i}F(\rx)h^{ij}(F(\rx), \rx)\p_{j}F(\rx)\right)<0, 
  \ (\hbox{ resp. }\leq 0).
  \]
  If $F$ is smooth then of course $\tilde{C}$ is space-like (resp. weakly space-like) iff all tangent vectors  at each point  of $\tilde{C}$ are  space-like (resp. space-like or null).

 Since $\tilde{\Sigma}$ is compact and $F$ Lipschitz, the Sobolev space $H^{1}(\tilde{C})$  and of course $L^{2}(\tilde{C})$ are well defined, for example by identifying $\tilde{C}$ with $\tilde{\Sigma}$ 
 and using the Riemannian metric $\tilde{h}_{ij}(0, \rx)d\rx^{i}d\rx^{j}$  on $\tilde{\Sigma}$ to equip $\tilde{C}$ with a density $d\nu_{\tilde{C}}$.  
 
 One also needs  the measure 
 \[
d \nu_{\tilde{C}}^{0}= (\beta^{-1}- h^{ij}\p_{i}\tilde{F}\p_{j}\tilde{F})d\nu_{\tilde{C}},
\]
which vanishes if $\tilde{C}$ is  a null hypersurface.  

We set now
\begin{equation}
\label{def-de-en}
\cE(\tilde{C})\defeq   H^{1}(\tilde{C})\oplus L^{2}(\tilde{C}, d \nu_{\tilde{C}}^{0}).
\end{equation}
Note that if $\tilde{C}$ is space-like (i.e. a Cauchy hypersurface), then $\cE(\tilde{C})= H^{1}(\tilde{C})\oplus L^{2}(\tilde{C})$. 

The result of \cite{Ho} is the following theorem:
\begin{theoreme}[\cite{Ho}]\label{22.1}
Let $\tilde{\Sigma}_{1}$ be any Cauchy hypersurface in $\tilde{M}$ and $\tilde{C}$ be weakly space-like of the form (\ref{formdeC}). Then the map
\[
\begin{aligned}
\tilde{T} :\, \cE(\tilde{\Sigma}_{1})\to & \ \cE(\tilde{C})\\
f\mapsto \,&\big( (\tilde{U}_{\tSig_{1}}f)\traa{\tilde{C}},  (\beta^{-1}\p_{t}\tilde{U}_{\tSig_{1}} f)\traa{ \tilde{C}}\big)
\end{aligned}
\]
is a homeomorphism.
\end{theoreme}
Note that if $\tilde{C}$ is characteristic, then $L^{2}(\tC, d \nu_{\tilde{C}}^{0})=\{0\}$ and $\cE(\tilde{C})=H^{1}(\tilde{C})$, so one obtains as a particular case the solvability of the characteristic Cauchy problem in energy spaces.  
\subsubsection{Embedding $M_{0}$ into $\tilde{M}$}\label{embedd}
We will use H\"{o}rmander's result recalled above to  solve a characteristic Cauchy problem in $M_{0}$, in an arbitrary neighborhood of $p$.   The first task is to locally embed $M$ into a spacetime $\tilde{M}$ as above.

We fix a Cauchy hypersurface $\Sigma$ to the future of $p$ and identify  $M$ with $\rr\times \Sigma$ with   
$g= - \beta(t,\rx)dt^{2} + h_{ij}(t,\rx)d\rx^{i}d\rx^{j}$.   We  set $\Sigma_{0}= \Sigma\cap M_{0}$ and
 fix   an  open, precompact set $U$ such that $J^{-}(\Sigma_{0})\cap J^{+}(p)\subset U$.

The following lemma shows that over $U$, $C$ can be parametrized by $\Sigma$.
\begin{lemma}\label{6.1}
 There exists a bounded, Lipschitz function $F$ defined on $\Sigma$ such that 
 \[
\overline{C}\cap U= \{(t, \rx) : \ t= F(\rx)\}\cap U.
\]
\end{lemma}
\proof The proof is given in Appendix \ref{proof6.1}. \qed 

\medskip

We next  embed $\Sigma_{0}$ into a smooth compact manifold 
$\tilde{\Sigma}$.    We consider the spacetime $\tilde{M}= \rr\times \tilde{\Sigma}$ and extend $F$ to a Lipschitz function $\tilde{F}$ on $\tilde{\Sigma}$, $g$ to a metric $\tilde{g}$ as in \ref{sec22.2.1}. We set
\[
\tilde{C}= \{t= \tilde{F}(\rx)\}\subset \tilde{M},
\]
and define:
\begin{equation}
\label{def-de-C_0}
C_{0}\defeq  \left(J^{-}(\Sigma_{0}; M)\cap C\right)\cup\{p\}.
\end{equation}
$C_{0}$ is an open subset of $\overline{C}$, with $\overline{C}_{0}$ compact in $M$ and
\begin{equation}
\label{prop-de-boundary}
\p \Sigma_{0}= \p C_{0}.
\end{equation}

We claim that we can choose the embedding $\Sigma_{0}\subset \tilde{\Sigma}$ and the extensions $\tilde{F}$ and $\tilde{g}$ so that: \begin{eqnarray}
\label{imp.1}
J^{-}(\tilde{\Sigma}\backslash \overline{\Sigma}_{0}; \tilde{M})\cap \bar{C}_{0}=\emptyset,\\[2mm]
\label{imp.2}\tilde{C}\hbox{ is weakly space-like in }\tilde{M}.
\end{eqnarray}
This is clearly possible by modifying $\Sigma$, $F$ and  $g$ only outside a large open set $U$,  and using that the embedding of $(M_{0}, g)$ into $(M, g)$ is causally compatible, see (\ref{e7.1}).

The situation is summarized in Fig. 2 below. Identification symbols (a single and double bar) are used to stress that $\tilde{\Sigma}$ is compact.

\begin{center}
\includegraphics[scale=0.8]{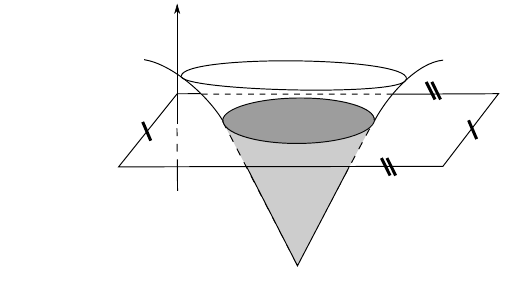}
\begin{picture}(0,0)\label{toto}
 \put(-90,10){$p$}
  \put(-100,65){$\Sigma_{0}$}
 \put(-100,35){$C_{0}$}
  \put(-132,55){$\tilde{\Sigma}$}
  \put(-65,60){$\p C_{0}$}
\put(-95, 95){$\tilde{C}$}
\put(-135, 100){$t$}
\end{picture}

Fig. 2: The modified cone $\tilde C$.
\end{center}
\subsubsection{Sobolev spaces}
We now recall some well-known facts about Sobolev spaces. If $\Omega$ is a relatively compact open set in a compact  manifold $X$ with smooth boundary $\p \Omega$, then $H^{1}_{0}(\Omega)$, defined as the closure of $\coinf(\Omega)$ in $H^{1}(\Omega)$ can also be characterized as $H^{1}_{0}(\Omega)= \{u\in H^{1}(\Omega) : \ u\tra{\p \Omega}=0\}$.
The restriction operator $r_{\Omega}: H^{1}(X)\to H^{1}(\Omega)$ is surjective from  $H^{1}_{\p\Omega}(X)= \{u\in H^{1}(X) : \ u\tra{\p \Omega}=0\}$ to $H^{1}_{0}(\Omega)$, with   right inverse $e_{\Omega}: H^{1}_{0}(\Omega)\to H^{1}_{\p\Omega}(X)$ equal to the extension by $0$ in $X\backslash \Omega$.

We set  $\cE_{0}(\Omega)\defeq H^{1}_{0}(\Omega)\oplus L^{2}(\Omega)$,  and 
$\cE_{\p\Omega}(X)= H^{1}_{\p \Omega}(X)\oplus L^{2}(X)$.  The operator $r_{\Omega}\oplus r_{\Omega}: \cE_{\p\Omega}(X)\to  \cE_{0}(\Omega)$ will still be denoted by $r_{\Omega}$, and  $e_{\Omega}\oplus e_{\Omega}: \cE_{0}(\Omega)\to \cE_{\p\Omega}(X)$ by $e_{\Omega}$. 

 We will use these facts for $\Omega= \Sigma_{0}, C_{0}$ and $X= \tilde{\Sigma}, \tilde{C}$. If $\Omega= C_{0}$, then we use the notation in (\ref{def-de-en}), i.e. $\cE_{0}(C_{0})= H^{1}_{0}(C_{0})\oplus \{0\}\sim H^{1}_{0}(C_{0})$, since $C_{0}$ is characteristic. 
\subsubsection{Characteristic Cauchy problem}
\textcolor{Red}{ In the theorem below, we denote by $U_{\Sigma_{0}}$ the operator $U_{\tilde{\Sigma}}\circ e_{\Sigma_{0}}$, i.e the Cauchy evolution operator (in $\tilde{M}$) for Cauchy data in $\cE_{0}(\Sigma_{0})$ (extended by $0$ in $\tilde{\Sigma}\backslash \Sigma_{0}$).}
 \begin{theoreme} \label{theothm}The map
\[
\begin{aligned}
T:\cE_{0}(\Sigma_{0})\to &\, \cE_{0}(C_{0})\\
f\mapsto &\, (U_{\Sigma_{0}}f)\traa{C_{0}}
\end{aligned}
\]
is a homeomorphism.
\end{theoreme}
\proof  We will prove the theorem by reducing ourselves to Thm. \ref{22.1}.  We first claim that
\begin{equation}
\label{e22.1}
T= r_{C_{0}}\circ \tilde{T}\circ e_{\Sigma_{0}}.
\end{equation}
In fact this follows from the fact that $e_{\Sigma_{0}}:\cE_{0}(\Sigma_{0})\to \cE(\tilde{\Sigma})$ is the extension by $0$.

 By Thm. \ref{22.1}, this implies that $T: \cE_{0}(\Sigma_{0})\to \cE(C_{0})$. Moreover by {\Red finite speed of propagation}, if $f\in \coinf(\Sigma_{0})\oplus \coinf(\Sigma_{0})$, then $Tf$ vanishes near $\p C_{0}$, hence $T$ maps  continuously $\cE_{0}(\Sigma_{0})$ into $\cE_{0}(C_{0})$. 

We next claim that $S= r_{\Sigma_{0}}\circ \tilde{T}^{-1}\circ e_{C_{0}}$ is a right inverse to $T$.  In fact let $g\in \cE_{0}(C_{0})$ and $\tilde{f}= \tilde{T}^{-1}\circ e_{C_{0}}g=(\tilde{f}^{0}, \tilde{f}^{1})\in\cE(\tilde{\Sigma})$. 
Since $\p \Sigma_{0}= \p C_{0}$ we have $\tilde{f}^{0}\tra{\p \Sigma_{0}}= g\tra{ \p C_{0}}=0$ hence $e_{\Sigma_{0}}\circ r_{\Sigma_{0}}\tilde{f}\in \cE(\tilde{\Sigma})$. Since $\tilde{f}-e_{\Sigma_{0}}\circ r_{\Sigma_{0}}\tilde{f}$ vanishes on $\overline{\Sigma}_{0}$, we obtain by (\ref{imp.1}) and {\Red finite speed of propagation} that
\[
r_{C_{0}}\circ \tilde{T}(\tilde{f}-e_{\Sigma_{0}}\circ r_{\Sigma_{0}}\tilde{f})=0,
\]
hence $T\circ S g= r_{C_{0}}\circ \tilde{T}\tilde{f}=r_{C_{0}}\circ e_{C_{0}}g=g$. This completes the proof of the theorem. 
\qed

\subsection{Sobolev space on the cone in null coordinates}
Let us set  \[
R: \cinf(C)\ni g\mapsto \beta^{-1}g(s, \theta)\in \cinf(\rr\times \bS).
\]
 The goal in this subsection is to describe more precisely the image of $H^{1}_{0}(C_{0})$ under $R$. 
 
 We will denote by $\tilde{C}_{0}\subset \rr\times \bS$ the image of  $C_{0}$ under the map $C\ni q\mapsto (s(q), \theta(q))$ where the coordinates $(s, \theta)$ are constructed in Lemma \ref{7.1}. Using that $\p C_{0}= \p \Sigma_{0}$ is space-like and included in $C$, we easily obtain from Lemma \ref{taz} that $\tilde{C}_{0}$ is of the form:
  \[
 \tilde{C}_{0}= \{(s, \theta)\in \rr\times \bS : \ s<s_{0}(\theta)\},
\]
for some smooth function $s_{0}$.  To simplify notation the measure $|m|^{\12}(\theta)d\theta$ on $\bS$ will be simply denoted by $d \theta$. We also set $r= \e^{s}$.
\begin{proposition}\label{propiprop} {\Red Assume $d= \dim M-1\geq 3$.} Then the image of $H^{1}_{0}(C_{0})$ under $R$ equals to the completion of $\coinf(\tilde{C}_{0})$ under the norm:
 \[
\| \psi\|_{1}\defeq   \bigg(\int_{\tilde{C}_{0}}(r^{-1} |\p_{s}\psi|^{2}+ r^{-1}|\p_{\theta}\psi|^{2} + r^{-1}| \psi|^{2})dsd\theta\bigg)^{\12}.
\]
We will denote this space by $\tilde{H}^{1}_{0}(\tilde{C}_{0})$.
\begin{remark}\label{remrem}
 Since $r\leq r_{0}$ on $C_{0}$, we see that $\tilde{H}^{1}_{0}(\tilde{C}_{0})$ injects continuously into $|D_{s}|^{-\12}L^{2}(\rr\times \bS)$.
 \end{remark}
\end{proposition}
\proof 
We recall that $(v,\psi)$ (see (\ref{e6.fourest})) are coordinates on $C$ such that the topology in $H_{0}^{1}(C_{0})$ is given by the norm
\[
\bigg(\int_{C_{0}}(|v|^{d-1} |\p_{v}g|^{2}+ | v|^{d-3}|\p_{\psi}g|^{2} + |v|^{d-1}| g|^{2})dvd\psi\bigg)^{\12}.
\]
Recall that we have set $r= \e^{s}$. A function $g\in H^{1}_{0}(C_{0})$ expressed in the coordinates $(s, \theta)$ or $(r, \theta)$ will be still denoted by $g$. Similarly the image of   $\tC_{0}$  under the map $(s, \theta)\mapsto (\e^{s}, \theta)$ will still be denoted by $\tC_{0}$.

 From Lemma \ref{7.1} (3)  and a routine computation, we see that  an equivalent norm on $H^{1}_{0}(C_{0})$ is:
\beq\label{enorm.1}
\bigg(\int_{\tC_{0}}(r^{d-1} |\p_{r}g|^{2}+ r^{d-3}|\p_{\theta}g|^{2} + r^{d-1}| g|^{2}drd\theta)\bigg)^{\12}.
\eeq
Since $d= \dim M-1\geq 3$,   the Hardy's inequality $-\Delta\geq C |x|^{-2}$ holds  on $L^{2}(\rr^{d})$. Considering $(r, \theta)$ as polar coordinates on $\rr^{d}$ we  obtain that:
\[
\int_{\tC_{0}}r^{d-1} |\p_{r}g|^{2}+ r^{d-3}|\p_{\theta}g|^{2} dr d\theta\geq C \int_{\tC_{0}}r^{d-3}|g|^{2} dr d \theta, \ g\in H^{1}_{0}(C_{0}).
\]  Therefore adding a term $r^{d-3}|g|^{2}$ under the integral sign in (\ref{enorm.1}) yields an equivalent norm on $H^{1}_{0}(C_{0})$.  Since $r$ is bounded on $\tC_{0}$ this term dominates the term $r^{d-1}|g|^{2}$ and we finally obtain that the topology of $H^{1}(C_{0})$ is given by the norm
\[
\bigg(\int_{\tC_{0}}(r^{d-1} |\p_{r}g|^{2}+ r^{d-3}|\p_{\theta}g|^{2} +\alpha r^{d-3}| g|^{2})drd\theta\bigg)^{\12},
\]
where the constant $\alpha>0$ can be chosen arbitrarily large.
Going  back to coordinates $(s, \theta)$ we obtain the norm
\beq\label{finalnorm}
\bigg(\int_{\tilde{C}_{0}}(r^{d-2} |\p_{s}g|^{2}+ r^{d-2}|\p_{\theta}g|^{2} + \alpha r^{d-2}| g|^{2})dsd\theta\bigg)^{\12}.
\eeq
  For two functions $m, n\in \cinf(\C_{0})$ we write $m\sim n$ if $m= r_{0}n$, for $r_{0}, r_{0}^{-1}\in S^{0}$, where the class $S^{0}$ is defined in Subsect. \ref{sec6.1.c}.  We have $\beta\sim r^{-(d-1)/2}$, hence
\beq\label{equiv}
 \p_{s}\beta, \ \p_{\theta}\beta\sim r^{- (d-1)/2}.
\eeq

Setting $\psi= Rg= \beta^{-1}g$ we have:
\[
\p_{s}g= \beta \p_{s}\psi+ (\p_{s}\beta) \psi, \ \p_{\theta}g= \beta\p_{\theta}\psi+ (\p_{\theta} \beta)\psi.
\]
Using then (\ref{equiv}), and choosing $\alpha\gg 1$ in (\ref{finalnorm}), we obtain that (\ref{finalnorm}) is equivalent to
\beq\label{finalnorm2}
\bigg(\int_{\tilde{C}_{0}}(r^{-1} |\p_{s}\psi|^{2}+ r^{-1}|\p_{\theta}\psi|^{2} + r^{-1}| \psi|^{2})dsd\theta\bigg)^{\12}.
\eeq
This completes the proof of the proposition. \qed 

\section{Change of null coordinates}\label{covar}\init
The map $\rho:\Sol(P_{0})\to \cH(\tC)$ introduced  in Def. \ref{10.1} depends on the choice of the null coordinates $(s, \theta)$ on $C$, i.e.  on the choice of the initial hypersurface $S$, used in Lemma \ref{7.1} to construct $(s, \theta)$. In this section we discuss 
how our class of Hadamard states depends on the above choice.

\subsection{New null coordinates}
We fix  a reference hypersurface $S$ in $C$,  yielding null coordinates $(s, \theta)$ near $C$ such that $g\tra{C}$ is given by (\ref{e6.03}) and $S= \{f=s=0\}$.

We  choose another  hypersurface $\tS$ transverse to $\nabla^{a}f$ in $C$, hence:
 \begin{equation}
\label{sua.e02}
\tS=\{f=0, \ s= b(\theta)\}, \hbox{ for some }b\in \cinf(\bS).
\end{equation}
Since $\nabla^{a}f\tra{C}= \p_{s}$, we obtain that the new coordinates $(\tilde{s}, \tilde{\theta})$ obtained from Lemma \ref{7.1} with $S$ replaced by $\tS$ are given by:
\begin{equation}
\label{sua.e03}
\tilde{\theta}= \theta, \ \tilde{s}(s, \theta)= s-b(\theta).
\end{equation}
 We have then  
 \[
 g\tra{C}= - 2 dfd\tilde{s}+ \tilde{h}_{ij}(\tilde{s}, \theta)d\theta^{i}d\theta^{j},
\]
and a standard computation shows that $|h|(\tilde{s}, \theta)= |h|(s, \theta)$, hence $\tilde{\beta}(\ts, \theta)= \beta(s, \theta)$. Denoting by $\tilde{\rho}$ the analog of $\rho$ in Def. \ref{10.1} for the new coordinates $(\ts, \theta)$ we have then\begin{equation}
\label{sua.05}
\tilde{\rho}\phi= U \rho\phi, \ \phi\in \Sol(P_{^{0}}),
\end{equation}
where:
\[
U:\begin{array}{l}
\cH(\tC)\to \cH(\tC)\\
g\mapsto Ug(s, \theta)= g(s+ b(\theta), \theta).
\end{array}
\]
The map $U$ is symplectic on $(\cH(\tC), \sigma_{C})$ and unitary on $L^{2}(\tC)$ with  $U^{*}D_{s}U = D_{s}$.

\begin{proposition}\label{sua.p1}
If $A\in \tPsi^{-\infty,p}(\tC)$ then $U AU^{-1}\in \tPsi^{-\infty, p}(\tC)$.
\end{proposition}
\begin{remark}
 Note that the above invariance property does not hold for the classes $\Psi^{m,p}(\tC)$,  since  for example the classes $\Psi^{m,p}(\rr\times \rr^{d-1})$ are not even preserved by linear changes of variables $(s, y)\mapsto (s+ Ay, y)$.
\end{remark}
\proof
We will use   the Beals criterion explained in Subsect. \ref{beals}, which implies that  $B\in \tPsi^{-\infty, p}(\tC)$ iff for any functions $g_{1}, \dots g_{n}\in \cinf(\bS)$ and smooth vector fields $X_{1},\dots, X_{m}$ on $\bS$  and for any $N\in \nn$, $k, k'\in \rr$ one has:
\beq\label{sua.e1}
\ad_{X_{1}}\cdots \ad_{X_{m}}\ad_{g_{1}}\cdots \ad_{g_{n}}B: H^{k, k'}(\tC)\to H^{k+N, k'-p+ n}(\tC).
\eeq
To simplify notation, we  rewrite (\ref{sua.e1}) as
\beq\label{sua.e3}
\ad_{\bar X}^{\alpha}\ad_{\bar g}^{\beta}B: H^{k, k'}(\tC)\to H^{k+N, k'+ p + |\beta|}(\tC),
\eeq
denoting by $\bar X$, resp. $\bar g$ an arbitrary $n-$uple of vector fields, resp. $m-$uple of functions.

If $g$ is a function on $\bS$, considered as a multiplication operator, and if $X$ is a vector field on $\bS$ we have:
\beq\label{sua.e2}
 U^{-1} gU= g, \ U^{-1} X U= X + (X\cdot db) \p_{s}, \ U^{-1}\p_{s}U = \p_{s}.
\eeq

Let  now $A\in \tPsi^{-\infty, p}(\tC)$.  For $\psi\in \cinf(\bS\times \bS)$, let us denote by $A_{\psi}$ the operator with distributional kernel $A(s_{1}, s_{2}, \theta_{1}, \theta_{2})\psi(\theta_{1}, \theta_{2})$. 
By the well-known properties of the pseudodifferential calculus on $\bS$   we know that if $\psi=1$ in some neighborhood of the diagonal, then $A- A_{\psi}\in \tPsi^{-\infty, -\infty}(\tC)$, or equivalently  maps $H^{k, k'}(\tC)$ into $H^{k+N, k'+N}(\tC)$ for any $k, k',N$.   Using (\ref{sua.e2}) this implies that $U(A- A_{\psi})U^{-1}$ has the same property, hence belongs to $\tPsi^{-\infty, -\infty}(\tC)$.

Therefore we can replace $A$ by $A_{\psi}$, and assume that the kernel of $A$ is supported in $\rr\times \rr\times \Omega$, where $\Omega$ is an arbitrarily small neighborhood of the diagonal in $\bS\times \bS$.  Introducing a smooth partition of unity $1= \sum_{1}^{M}\chi_{i}$ on $\bS$, we see that we can replace $A$ by $ \chi A \chi$, where $\chi\in \cinf(\bS)$ is supported in a small neighborhood of a point $\theta_{0}\in \bS$. We pick local coordinates $\theta_{1}, \dots,\theta_{d-1}$ near $\theta_{0}$ and  rewrite (\ref{sua.e3}) as:
\begin{equation}
\label{sua.e4}
\langle \p_{s}\rangle^{k+N}\langle \p_{\theta}\rangle^{k'-p+|\beta|} \ad_{\bar X}^{\alpha}\ad_{\bar g}^{\beta}A \langle \p_{s}\rangle^{-k}\langle \p_{\theta}\rangle^{-k'}\in B(L^{2}(\tC)).
\end{equation}
We set now $A'=UA U^{-1}$. Note first that if the kernel of $A$ is supported in $\rr\times\rr\times \Omega$, then so is   the kernel of $A'$, hence by the above discussion it suffices to check  that $A'$ satisfies (\ref{sua.e4}).  Let us set $U^{-1}XU=X'$ if $X$ is a vector field on $\bS$, and in particular $\p_{\theta}'= U^{-1}\p_{\theta}U= \p_{\theta}+\p_{\theta}b \p_{s}$.
Then an easy computation yields:
\beq\label{sua.e6}
\begin{aligned}
&\langle \p_{s}\rangle^{k+N}\langle \p_{\theta}\rangle^{k'-p+|\beta|} \ad_{\bar X}^{\alpha}\ad_{\bar g}^{\beta}UAU^{-1} \langle \p_{s}\rangle^{-k}\langle \p_{\theta}\rangle^{-k'}\\[2mm]
&=U\langle \p_{s}\rangle^{k+N}\langle \p_{\theta}'\rangle^{k'-p+|\beta|} \ad_{\bar X'}^{\alpha}\ad_{\bar g}^{\beta}A \langle \p_{s}\rangle^{-k}\langle \p_{\theta}'\rangle^{-k'}U^{-1}.
\end{aligned}
\eeq
Using (\ref{sua.e2}) and the fact that $A\in \tPsi^{-\infty, p}(\tC)$, we obtain that 
\[
\ad_{\bar X'}^{\alpha}\ad_{\bar g}^{\beta}A\in \tPsi^{-\infty, p-|\beta|}(\tC),
\]
\[
\langle \p_{s}\rangle^{N}\langle \p_{\theta}\rangle^{k'-p+ |\beta|}\ad_{\bar X'}^{\alpha}\ad_{\bar g}^{\beta}A\langle \p_{s}\rangle^{N} \langle \p_{\theta}\rangle^{-k'}\in B(L^{2}(\tC)),
\]
for any $N\in \nn$.
It follows that the l.h.s. of (\ref{sua.e6}) belongs to $B(L^{2}(\tC))$ if for any $s\in \rr$ there exists $N\in \nn$ such that:
\begin{equation}
\label{sua.e7}
\langle \p_{s}\rangle^{-N} \langle \p_{\theta}
'\rangle^{s}\langle \p_{\theta}\rangle^{-s}, \ \langle \p_{s}\rangle^{-N} \langle \p_{\theta}\rangle^{s}\langle \p_{\theta}'\rangle^{-s}\in B(L^{2}(\tC)).
\end{equation}
Let  us now prove (\ref{sua.e7}). The first statement of (\ref{sua.e7}) is easy to check for $s\in \nn$, using that $\p_{\theta}'= \p_{\theta}+ \p_{\theta}b\p_{s}$. Conjugation by $U$ gives the second statement for $s\in \nn$. By duality and interpolation we obtain then (\ref{sua.e7}) for arbitrary $s$, which completes the proof of the proposition. \qed

\medskip

From Prop. \ref{sua.p1} and the fact that $U^{*}D_{s}U= D_{s}$, we obtain immediately  the following result.

\begin{proposition}
\label{sua.p2}
The classes of Hadamard states obtained in Thms. \ref{thm:constr}, \ref{thm6}  are independent on the choice of the null coordinates $(s, \theta)$. 
\end{proposition}

 \appendix
\section{}\init\label{secapp1}
\subsection{Stokes  formula}\label{secapp1.1}
Let $(M, g)$ an orientable, oriented  pseudo-Riemannian manifold of dimension $n$. We denote by $d{\rm Vol}_{g}\in \forms^{n}(M)$ the associated volume form and by $d\mu_{g}= | d{\rm Vol}_{g}|$  the associated density. 

Let $\Sigma\subset M$ a smooth submanifold of codimension $1$ and $\iota: \Sigma\to M$ the natural injection, which induces 
$\iota^{*}: \forms(M)\to \forms(\Sigma)$. From the  orientation of $M$ and  a continuous transverse vector field $v\in T_{\Sigma}M$  we obtain an induced orientation of $\Sigma$. If $\Sigma\subset \p U$ for an open set $U\subset M$ {\Red with piecewise smooth boundary $\p U$}, we choose $v$ pointing outwards.

If $\omega\in \forms^{n}(M)$ and $X\in T M$, then  $X\lrcorner\,\omega\in \forms^{n-1}(M)$ and one sets:
\[
\iota_{X}^{*}\omega\defeq  \iota^{*}(X\lrcorner\,\omega)\in \forms^{n-1}(\Sigma).
\]
Similarly if $\mu= |\omega|$ is a density on $M$, we set $\iota_{X}^{*}\mu\defeq  | \iota_{X}\omega|$, which is a density on $\Sigma$. 

If $\nabla_{a}$ is the Levi-Civita connection associated to $g$ then:
\[
\nabla_{a}X^{a}d{\rm Vol}_{g}= d (X\lrcorner\,d{\rm Vol}_{g}),
\]
which applying Stokes formula:
\begin{equation}
\label{e3.1}
\int_{U}d\omega= \int_{\p U}\iota^{*}\omega, \ \omega\in \forms^{n-1}(M),
\end{equation}
to $\omega= \iota_{X}^{*} d{\rm Vol}_{g}$, yields:
\begin{equation}
\label{e.app2}
\int_{U}\nabla_{a}X^{a} d{\rm Vol}_{g}= \int_{\p U} \iota^{*}_{X} d{\rm Vol}_{g}.
\end{equation}

\subsubsection{Non-characteristic boundaries}
 Assume first  $\Sigma\subset \p U$ is {\em non characteristic}, that is  the one-dimensional space:
\[
T_{x}(\Sigma)^{\rm ann}\subset T_{x}M^{*}
\]
is not {\em null} (the superscript $^{\rm ann}$ denotes the annihilator).  It follows that the metric $h\defeq  \iota^{*}g$ on $\Sigma$ is non-degenerate (in the Lorentzian case, one typically assume that $\Sigma$ is space-like, then $h=\iota^{*}g$ is Riemannian).  Let $n\in T_{\Sigma}M$ be the unit, outward pointing normal vector field to $\sigma$. 
Then : 
\beq\label{e.app1}
d{\rm Vol}_{h}= \iota_{n}^{*} d{\rm Vol}_{g}, \ \iota^{*}_{X} d{\rm Vol}_{g}= X^{a}n_{a} d{\rm Vol}_{h},
\eeq
hence 
\[
\int_{\Sigma} \iota^{*}_{X} d{\rm Vol}_{g}\defeq  \int_{\Sigma} X^{a}n_{a} d\sigma_{h}.
\]
If all of $\p U$ is non-characteristic, then from (\ref{e.app2}) 
 we obtain Gauss' formula:
\beq\label{e3.2}
\int_{U}\nabla_{a}X^{a} d\mu_{g}= \int_{\Sigma} X^{a}n_{a} d\sigma_{h},
\eeq
where $d\sigma_{h}= | d{\rm Vol}_{h}|$.
 
\subsubsection{Characteristic boundaries}
Assume now that  $\Sigma$  is characteristic. Then there is no normal vector  field anymore. To express the r.h.s. of (\ref{e.app2}), one  chooses a  defining function function $f$ for $\Sigma$, i.e. such that   $f=0, df\neq 0$ on $\Sigma$, and complete $f$ with coordinates $y^{1}, \dots, y^{n-1}$ such that $df\wed dy^{1}\wed\cdots \wed dy^{n-1}$ is positively oriented. Then computing in the coordinates $f, y^{1}, \dots, y^{n-1}$ one sees that
\[
\iota_{X}^{*} d{\rm Vol}_{g}= X^{a}\nabla_{a}f |g|^{\12} dy^{1}\wedge\cdots \wedge dy^{n-1},
\]
hence:

\begin{equation}
\label{e3.3}
\int_{\Sigma} \iota^{*}_{X} d{\rm Vol}_{g}=\int_{\Sigma}X_{a}\nabla^{a}f|g|^{\12}  dy^{1}\wed\cdots \wed dy^{n-1}
\end{equation}
In the general case we can for example split $\p U$  as $\Sigma_{1}\cup \Sigma_{2}$, where $\Sigma_{1}$ is non-characteristic, $\Sigma_{2}$ is characteristic and obtain:
\begin{equation}
\label{e.app3}
\int_{U}\nabla_{a}X^{a} d\mu_{g}=  \int_{\Sigma_{1}} X^{a}n_{a} d\sigma_{h}+  \int_{\Sigma_{2}}X_{a}\nabla^{a}f|g|^{\12}  dy^{1}\wed\cdots \wed dy^{n-1}.
\end{equation}

\subsection{Conformal transformations}\label{ss:confo}

In this section we briefly discuss conformal transformations of a globally hyperbolic spacetime $(M,g)$. Let $\omega\in C^\infty(M)$ be strictly positive and consider the conformally related metric 
\[
g'=\omega^2 g.
\]
Set $P= -\nabla^{a}\nabla_{a}+ \frac{n-2}{4(n-1)}R$, {\Red where $R$ is the scalar curvature}. For this special choice of the lower order terms, the conformal transformation $g\to g'$ amounts to
\[
P'= \omega^{-n/2-1} P \omega^{n/2-1}.
\]
This entails that the causal propagators are related by $E'=\omega^{-n/2+1} E \omega^{n/2+1}$. One concludes that multiplication by $\omega^{-n/2+1}$ induces a symplectic map
\beq\label{confss}
(\Sol(P),\sigma)\xrightarrow{ \ \omega^{-n/2+1} \ }(\Sol(P'),\sigma'),
\eeq
where $\sigma,\sigma'$ are defined as in (\ref{e7.5}) using the respective volume densities.

We apply this discussion to $(M_0,g)$ and the conformally related spacetime with metric $g'=\omega^2 g$. In the setting of Sect. \ref{sec10.2}, there is a monomorphism of symplectic spaces
\[
(\Sol(P_0),\sigma_0)\xrightarrow{\ \rho \ }(\cH(\tilde{C}),\sigma_C).
\] 
By (\ref{confss}) we also have a monomorphism
\[
(\Sol(P'_0),\sigma'_0)\xrightarrow{\ \rho \circ\omega^{n/2-1} \ }(\cH(\tilde{C}),\sigma_C).
\] 
Therefore, one can construct states for the conformally related spacetime using the bulk-to-boundary correspondence with a modified trace map $\rho'=\rho\circ\omega^{n/2-1}$.


\subsection{Proof of Lemma \ref{taz}}\label{proof-of-lemid}
\def\bx{\bar{x}}
We fix a point $q\in C$ and  complete the coordinate $x^{0}= f$ by local coordinates $\bar{x}= (x^{1},\dots, x^{d})$ near $q$. The functions $s, \theta_{k}$ defined on $C$ are denoted by $s(\bx), \theta_{k}(\bx)$, since $\bx$ are local coordinates on $C$. We denote by $h(\bx)$ the  restriction of $g^{-1}$ to $T^{*}C$. Note that the fact that  $C$ is null implies that $g^{00}(0, \bx)\equiv 0$ and that from Lemma \ref{7.1} we have:
\beq\label{vic}
g^{i0}(\bx)\p_{i}s(\bx)= -1, \ g^{i0}(\bx)\p_{i}\theta_{k}(\bx)= 0.
\eeq
 If $X$ is a null vector, orthogonal to $C\cap \{s(\bx)= s(q)\}$  and transverse to $C$, 
 we obtain that \[
gX= \lambda(\12 \nabla_{i}s \nabla^{i}s, \nabla_{i}s ), \ \lambda\in \rr.
\]
Let us denote for the moment by $\tilde{s}$, $\tilde{\theta}_{k}$ the extensions of $s, \theta_{k}$ outside $C$, which are constant along the flow of $X$. We obtain that on $C$:
\[
 d \tilde{s}= (\12  ds\cdot h ds, ds), \ d \tilde{\theta}_{k}= (ds\cdot hd \theta_{k}, d\theta_{k}).
\]
Using also $df=(1, 0, \dots, 0)$ and (\ref{vic}), a routine computation leads to the following identities on $C$: 
\[
\begin{array}{l}
df\cdot  g^{-1}df= d\tilde{s}\cdot  g^{-1}d\tilde{s}= df \cdot g^{-1}d\tilde{\theta}_{k}= d\tilde{s}\cdot g^{-1}d\tilde{\theta}_{k}= 0, \\[2mm]
df\cdot g^{-1}d\tilde{s}= d\tilde{s}\cdot g^{-1} df=-1,\\[2mm]
d\tilde{\theta}_{k}\cdot g^{-1} d\tilde{\theta}_{l}= \p_{i}\theta_{k}h^{ij}\p_{j}\theta_{l}.
\end{array}
\]
This implies that $g$ is of the form (\ref{e6.03}) on $C$. \qed

\subsection{Proof of Lemma \ref{7.1}.}\label{fred-crook}

\def\by{\overline{y}}
Since  $(y^{0}, \overline{y})$ are normal coordinates, we have:
\beq\label{e7.2}
 g\tra{C}= -dv d w+ \frac{1}{4}v^{2} m_{ij}(\psi)d \psi^{i}d \psi^{j}+ v^{2}g_{1}, 
\eeq
where $m_{ij}(\psi)d \psi^{i}d \psi^{j}$ is the standard Riemannian metric on $ \bS$ and $g_{1}$ is a smooth pseudo-Riemannian  metric in the arguments $ dv$, $dw$ and $v d\psi^{i}$.

We start by expressing $f$ in the normal coordinates $(y^{0}, \by)$. By Malgrange's preparation theorem \cite[Thm. 7.5.6]{H1}  one can write
\[
f(y^{0}, \by)= m(y^{0}, \by)((y^{0})^{2}- |\by|^{2})+ a(\by)y^{0}+ b(\by),
\]
for $m$, resp. $a,b\in C^{\infty}$ near $(0, 0)$, resp. near $0$. Since $C\subset f^{-1}(\{0\})$, we obtain that $b(\by)= a(\by)|\by|$,  and since $b\in \cinf(\rr^{d})$, necessarily $a\in O(|\by|^{\infty})$. Moreover from the Hessian of $f$ at $p$ we obtain that $m(0, 0)=1$.

Going to coordinates $(v, w, \psi)$, we obtain:
\[
f(v, w, \psi)= m(v, w, \psi)vw+ wa(v, w, \psi),
\]
for $a\in O(|w-v|^{\infty})$. Using also that $m(0, 0, \psi)=1$, it follows that:
\[
\p_{v}f(v, 0, \psi)= \p_{\psi^{i}}f(v, 0 , \psi)=0, \ \p_{w}f(v, 0 , \psi)= v+  r(v, \psi),
\]
for $r\in O(|v|^{2})$.   Using (\ref{e7.2}) to express $(g^{-1})\traa{C}$ we obtain after an easy computation that:
\beq\label{fred}
\nabla^{a}f= -2v\left((1+ va^{0}(v, \psi))\p_{v}+ va^{i}(v, \psi)\p_{\psi^{i}}\right),
\eeq
where $a^{0}, a^{i}$ are smooth, bounded functions near $v=0$.

Let us now prove (1). Using (\ref{fred})  we  obtain the equation near $p$:
\[
(v+ v^{2}a^{0}(v, \psi))\p_{v}s + v^{2}a^{i}(v, \psi)\p_{\psi^{i}}s=\12,
\]
for smooth functions $a^{0}$, $a^{i}$. We set $s=\12\ln( vh(v, \psi))$ and obtain after an elementary computation
\[
 (1+ v a^{0})\p_{v}h+ a^{0} h+  va^{i}(v, \psi) \p_{\psi^{i}}h=0,
\]
which we can uniquely solve on $[-\epsilon_{1}, \epsilon_{1}]\times\bS$ by fixing $h(0, \psi)$. We may fix $h(0, \psi)>0$ to ensure that $s(\epsilon_{0}, \psi)=0$. We obtain $s= \12\ln(v)+ \12\ln h(v, \psi)$ for $h\in \cinf([-\epsilon_{1}, \epsilon_{1}]\times\bS)$, $h>0$.

It remains to extend $s$ globally to $C$.  To do this it suffices to check that  for any $q\in C$, the integral curve of 
$\nabla^{a}f$  through $q$ crosses $S$ at one and only one point.  By \cite[Corollary to Thm. 8.1.2]{W}  we know that $q$ can be joined to $p$ by a null geodesic $\gamma$. Locally a null geodesic on $C$ is, modulo reparametrization, an integral curve of $\nabla^{a}f$. Since $\nabla^{a}f$ is complete, the whole $\gamma\backslash\{p\}$ is an integral curve of $\nabla^{a}f$.  Hence the integral curve of 
$\nabla^{a}f$ through $q$ crosses $S$. Choosing $\epsilon_{0}$ in (\ref{def-de-surf}) small enough, we can ensure that $\nabla^{a}f\nabla_{a}v>0$ on $S$, hence the integral curve through $q$ crosses $S$ at only one point. 
 We can hence extend $s$ globally to $C$, as a $C^{\infty}$ function.

  The proof of (2) is similar. We obtain  the equation near $p$:
  \[
(v+ v^{2}a^{0}(v, \psi))\p_{v}\theta^{j} + v^{2}a^{i}(v, \psi)\p_{\psi^{i}}\theta^{j}=0,
\]
or equivalently:
\[
(1+ v a^{0}(v, \psi))\p_{v}\theta^{j} + v a^{i}(v, \psi)\p_{\psi^{i}}\theta^{j}=0,
\]
which we can solve in $]-\epsilon_{1}, \epsilon_{1}[\times\bS$ by imposing $\theta^{j}(\epsilon_{0}, \psi)= \psi^{j}$.  The estimate (3) on $\theta^{j}$ is immediate.  We extend $\theta^{j}$ to all of $C$ by the same argument as before.
  \qed

\subsection{Proof of Lemma \ref{schulz.bis}}\label{A2}
We use the characterization of the wavefront set of kernels using oscillatory  test functions, which we now recall:

let  $(\tilde s, \tilde y)\in C$ and $\lambda\geq 1$. We set   for $(\sigma, \eta)\in \rr\times \rr^{d-1}$:
\beq\label{e.norm}
v_{\sigma, \lambda}(\cdot)= \chi(\cdot)\e^{\i \lambda \langle \cdot,  \sigma\rangle }\in \coinf(\rr), \ w_{\eta, \lambda}(\cdot)= \psi(\cdot)\e^{\i \lambda\langle \cdot,  \eta\rangle}\in \cinf(\bS),
\eeq
where $\chi\in \coinf(\rr)$, resp. $\psi\in \cinf(\bS)$ are supported near $\tilde s$, resp.  $\tilde y$.  We set $u_{(\sigma, \eta), \lambda}= v_{\sigma, \lambda}\otimes w_{\eta, \lambda}$. Note that if $V$ resp. $W$ are small neighborhoods of $\tilde \sigma\in \rr$, resp. $\tilde  \eta\in \rr^{d-1}$ then  for $n_{+}= \max(n, 0)$, we have uniformly on $U= V\times W$:
\begin{equation}
\label{e.s0}
\| u_{(\sigma, \eta), \lambda}\|_{k, k'}\in 
\begin{cases}
O(\nlam^{k_{+}+ k'_{+}})\\
O(\nlam^{k+ k'_{+}})\hbox{ if }\sigma_{0}\neq 0,\\
O(\nlam^{k_{+}+ k'})\hbox{ if }\eta_{0}\neq 0.
\end{cases}
\end{equation}
Let now $\tilde Y_{1}, \tilde Y_{2}\in T^{*}C$.   Then $(\tilde Y_{1}, \tilde Y_{2})\not\in \WF(a)'$ if there exists  cutoff functions $\chi_{i}$, resp. $\psi_{i}$ with $\chi_{i}(\tilde s_{i}), \psi_{i}(\tilde y_{i})\neq 0$ and neighborhoods $U_{i}= V_{i}\times W_{i}$ of $(\tilde \sigma_{i}, \tilde \eta_{i})$ such that:
\begin{equation}
\label{e.s1}
(u_{(\sigma_{1}, \eta_{1}), \lambda}| a u_{(\sigma_{2},\eta_{2}) , \lambda})_{L^{2}(C)}\in O(\nlam^{-\infty}), \hbox{uniformly for }(\sigma_{i}, \eta_{i})\in U_{i}. 
\end{equation}
We first prove (1). Let $a\in B^{-\infty}\Psi^{p_{2}}(C)$ and $\tilde Y_{1}, \tilde Y_{2}\in T^{*}C$ such that $\tilde{\sigma}_{1}\neq 0$ or $\tilde{\sigma}_{2}\neq 0$.
Then  (\ref{e.s1}) follows from (\ref{e.s0}) and the fact that $a: H^{k_{1}, k_{2}}\to H^{k_{1}+m, k_{2}+ p_{2}}$ for any $m\geq 0$.

We now prove (2). If $a\in \Psi^{p_{1}, p_{2}}(C)$ the statement follows from Lemma \ref{schulz}. It remains to consider the case $a\in B^{-\infty}\Psi^{p_{2}}(C)$, and 
 to prove that (\ref{e.s1}) holds if $(\tilde \sigma_{1},\tilde  \eta_{1})=(0, 0)$ and $(\tilde \sigma_{2}, \tilde \eta_{2})\neq 0$ or vice versa.   If $\tilde{\sigma}_{1}\neq 0$ or $\tilde{\sigma}_{2}\neq 0$  we have already proved (\ref{e.s1}).
 
   Assume now that $\tilde \eta_{1}=0$ and $\tilde \eta_{2}\neq 0$, the other case being similar. Then we can find  cutoff functions $g_{i}\in \coinf(\rr^{d-1})$ supported near $\tilde \eta_{i}$, with disjoint supports such that $(1- g_{i}(\lambda^{-1}D_{y}))u_{(\sigma_{i}, \eta_{i}), \lambda}\in O(\lambda^{-\infty})$ in all $H^{k, k'}$, uniformly for $(\sigma_{i}, \eta_{i})\in U$. It  follows that 
 \[
 \begin{aligned}
&(u_{(\sigma_{1}, \eta_{1}), \lambda}| a u_{(\sigma_{2},\eta_{2}) , \lambda})_{L^{2}(C)}\\[2mm]
&= (u_{(\sigma_{1}, \eta_{1}), \lambda}| g_{1}(\lambda^{-1} D_{y})a g_{2}(\lambda^{-1}D_{y}) u_{(\sigma_{2},\eta_{2}) , \lambda})_{L^{2}(C)}+ O(\nlam^{-\infty}), 
\end{aligned}
\]
uniformly  for $(\sigma_{i}, \eta_{i})\in U_{i}$.  By pseudodifferential calculus on $\bS$, we know that $g_{1}(\lambda^{-1}D_{y})ag_{2}(\lambda^{-1}D_{y})\in O(\nlam^{-\infty})$ in $B(H^{k, k'})$ for any $k, k'\in \rr$. Combined with (\ref{e.s0}), we obtain (\ref{e.s1}) also if $\tilde \eta_{1}=0$, $\tilde \eta_{2}\neq 0$. This completes the proof of the lemma. \qed
\subsection{Proof of Lemma \ref{6.1}}\label{proof6.1}
Set $\gamma_{\rx}= \{(s,\rx):  s\leq 0\}$, $\rx\in \Sigma$.
To prove that $\overline{C}$ is the graph of a function $F$ over $\Sigma$ we have to show that for each $\rx\in \Sigma$, $\gamma_{\rx}$ intersects $\overline{C}$ at one and only one point. Then we have:
\[
F(\rx)= \inf\{s\leq 0: (s, \rx)\in I^{+}(p)\}.
\]
If $F(\rx)= -\infty$ then $\gamma_{\rx}\subset I^{+}(p)\cap J^{-}((0,\rx))\subset J^{+}(p)\cap J^{-}((0, \rx))$. This last set is compact by global hyperbolicity, which is a contradiction.  Hence $\gamma_{\rx}$ intersects $\overline{C}$. Moreover if $(t_{1}, \rx)\in \overline{C}$, then $(s,\rx)\in J^{-}(p)$ for all $ t_{1}\leq s \leq 0$. This shows that $\gamma_{\rx}$ intersects $\overline{C}$ at only one point, hence the function $F$ is well defined, and bounded.

 Let $(T^{0}, \rx^{0})$ the coordinates of $p$. For $\rx\neq \rx^{0}$, $C$ is smooth near $(F(\rx), \rx)$ and $\p_{t}$ is transverse to $C$. By the implicit function theorem this implies that $F$ is smooth near $\rx$. Moreover if $K_{1}\subset\Sigma$ is a compact set  then $d F$ is uniformly  bounded on $K_{1}\backslash\{x^{0}\}$. To prove this is suffices to introduce normal coordinates at $p$ such that near $p$, $C$ becomes a neighborhood of the tip of the  flat lightcone.
\qed

\subsection*{Acknowledgments} The work of M.W.\ was partially supported by the FMJH (French Government Program:  ANR-10-CAMP-0151-02).

\end{document}